\documentclass[twocolumn,showpacs,preprintnumbers,amsmath,amssymb,floatfix]{revtex4}
\usepackage[caption=false]{subfig}
\usepackage{graphicx}

\begin{document}

\title{Planar domain walls in black hole spacetimes}
\author{Filip Ficek}
 \email{filip.ficek@uj.edu.pl}
\author{Patryk Mach}
\email{patryk.mach@uj.edu.pl}
\affiliation{
Institute of Physics, Jagiellonian University, \L{}ojasiewicza 11, 30-348 Krak\'{o}w, Poland
}

\date{\today}

\begin{abstract}
We investigate the behaviour of low-mass, planar domain walls in the so-called $\phi^4$ model of the scalar field on the Schwarzschild and Kerr backgrounds. We focus on a transit of a domain wall through a black hole and solve numerically the equations of motion for a range of parameters of the domain wall and the black hole. We observe a behavior resembling an occurrence of ringing modes. Perturbations of domain walls vanish during latter evolution, suggesting their stability against a passage through the black hole. The results obtained for Kerr and Reissner-Nordstr\"{o}m black holes are also compared.
\end{abstract}


\maketitle

\section{Introduction}
Topological defects are present in many Beyond Standard Model theories and cosmological models \cite{Vil}. In the cosmological context, several authors suggested that they could be responsible for creating the large scale structure \cite{Hil88, Sik82}, driving the inflation \cite{Vil94, Lin90}, or acting as a Dark Matter or the Dark Energy \cite{Hol83, Buc90, Fri03}. In this paper we investigate defects occurring in models with a disconnected vacuum manifold --- domain walls. In particular, we study a passage of a planar domain wall through a Kerr black hole.

Solutions representing static domain walls near black holes were investigated for the Schwarzschild \cite{Mor00, Mor03}, Reissner-Nordstr\"{o}m \cite{Mod04}, and Schwarzschild-anti-de Sitter \cite{Mod06} spacetimes. An evolutionary scenario was considered by Flachi, Pujol\`{a}s, Sasaki and Tanaka, who studied an escape of a domain wall from the vicinity of a higher dimensional Schwarzschild black hole \cite{Fla06}. Here we focus on the whole process of a transit of a scalar field domain wall through a black hole.

From the technical point of view, we study the dynamics of a nonlinear scalar field in the Kerr geometry. This touches upon the stability of the Kerr spacetime \cite{Daf09, Daf11} --- probably one of the most important open problems in mathematical General Relativity \cite{Pre73, Hod12, Dol07}. On the other hand, because in our analysis the Kerr background is fixed, our results are rather relevant to the understanding of the stability of domain walls. Our work was also motivated by the current interest in the experimental detection of domain walls, using both astrophysical observations \cite{Sta14, Sta14b} and earthbound experiments \cite{Pos13, Der14, Pus13}. In the context of this research, it is interesting to learn how domain walls behave near astrophysical objects.

In the following sections we report results of simulations of a transit of a domain wall through a black hole in an axially-symmetric setting. We investigated the impact of the spin of the black hole on such a process. We show that, in spite of a formation of an additional structure resembling ringing modes, the domain wall is stable under its transit through the black hole --- it returns to its initial shape. We also compare results obtained for transits through Kerr and Reissner-Nordstr\"{o}m black holes.

Throughout this paper we consider 4-dimensional metrics with the signature $(-,+$, $+,+)$. We use natural units ($c=G=1$). Spacetime and spatial coordinates are labeled with Greek ($\mu$, $\nu$, \dots) and Latin ($i$, $j$, \dots) indices, respectively. We also use standard Einstein summation convention. The time coordinate is labeled with $t$ or index $0$.

\section{Domain walls in the Minkowski spacetime}

Consider a real scalar field in a $d+1$ dimensional flat Minkowski spacetime with a Lagrangian density
\begin{equation} 
\mathcal{L} = -\frac{1}{2}\partial_\mu \phi \partial^\mu \phi - V(\phi).
\label{eqn:LagFlat}
\end{equation}
The corresponding equation of motion (the Euler-Lagrange equation) reads
\begin{equation} 
\Box\phi-V'(\phi)=0,
\label{eqn:EomFlat}
\end{equation}
where $\Box=\partial_\mu\partial^\mu$ is the d'Alembert operator. The energy density (the $T_{00}$ component of the energy-density tensor $T_{\mu \nu}$) can be written as
\begin{equation} 
T_{00}=\frac{1}{2}\left(\partial_t \phi\right)^2+\frac{1}{2}\sum_{i=1}^d \left(\partial_i \phi\right)^2+V(\phi).
\label{eqn:EnergyDensityFlat}
\end{equation}
The total energy of the field configuration can be obtained by integrating $T_{00}$ over the $t = \mathrm{const}$ hypersurface.

In this paper we work with the so-called $\phi^4$ model defined by the field potential $V(\phi)$ of the form
\begin{equation} 
V(\phi)=\frac{\lambda}{4}\left(\phi^2-\eta^2\right)^2,
\label{eqn:Potential4}
\end{equation}
where $\lambda$ and $\eta$ are constant \cite{Aro, Vac}. We will refer to $\lambda$ and $\eta$ as the coupling constant and the vacuum expectation value, respectively.

It is a handbook knowledge that the above definition leads to an occurrence of the spontaneous symmetry breaking. The Lagrangian density $\mathcal{L}$ with the potential given by Eq.\ (\ref{eqn:Potential4}) has the $\mathbb{Z}_2$ symmetry ($\phi \to -\phi$) and two distinct ground states: $\phi\equiv\eta$ and $\phi\equiv-\eta$. This $\mathbb{Z}_2$ symmetry is broken by a choice of a particular ground state ($\phi\equiv\eta$ or $\phi\equiv-\eta$). It is also well known that the model of the scalar field with the potential (\ref{eqn:Potential4}) admits domains and domain walls. Consider $\textbf{x}_1$ and $\textbf{x}_2$ --- two distinct points in space such that at some instant of time $t$ one has $\phi(t,\textbf{x}_1)=\eta$ and $\phi(t,\textbf{x}_2)=-\eta$. At both these points the field attains a minimum of the potential. From the continuity of $\phi$, somewhere between these two points, we must have $\phi=0$, i.e., the field attains its local maximum of potential. The set of points $\textbf{x}$ such that $\phi(t,\textbf{x})=0$ together with its neighbourhood is called a domain wall. The regions where $\phi(t,\textbf{x})\approx \eta$ and $\phi(t,\textbf{x})\approx -\eta$ are understood as distinct domains.

Consider a $1 + 1$ dimensional spacetime with coordinates $(t,z)$, and the scalar field $\phi$ with potential (\ref{eqn:Potential4}). The equation of motion has the form
\begin{equation} 
-\partial^2_t\phi+\partial^2_z\phi=\lambda\phi^3-\lambda\eta^2\phi.
\label{eqn:EomFlat1}
\end{equation}
An example of a non-trivial, static solution of this equation with the boundary conditions $\lim_{z\to-\infty}\phi(z)=-\eta$ and $\lim_{z\to+\infty}\phi(z)=\eta$ is the so-called kink \cite{Vac, Aro},
\begin{equation} 
\phi(z)=\eta\tanh\left(\eta\sqrt{\frac{\lambda}{2}}(z-z_0)\right).
\label{eqn:SolMink1}
\end{equation}
Here $z_0$ is a constant. Non-static solutions can be obtained by applying a Lorentz boost to  Eq.\ (\ref{eqn:SolMink1}). One gets
\begin{equation} 
\phi(t,z)=\eta\tanh\left(\eta\sqrt{\frac{\lambda}{2}}\frac{z-z_0-v t}{\sqrt{1-v^2}}\right),
\label{eqn:SolMink2}
\end{equation}
where $v$ is the boost velocity. It can be easily checked that (\ref{eqn:SolMink2}) satisfies Eq.\ (\ref{eqn:EomFlat1}). The total energy of this solution is finite. It reads
\begin{equation} 
E_{kink}=\frac{1}{\sqrt{1-v^2}}\frac{4}{3}\sqrt{\frac{\lambda}{2}}\eta^3.
\label{eqn:Energy}
\end{equation}

The $1+1$ dimensional solution can be trivially generalised to $d+1$ dimensions as
\begin{equation} 
\phi(t,x_1,x_2,...,x_d)=\eta\tanh\left(\eta\sqrt{\frac{\lambda}{2}}\frac{x_1-z_0-v t}{\sqrt{1-v^2}}\right).
\label{eqn:SolMinkd}
\end{equation}
The above class of solutions is known as planar domain walls.

\begin{figure}
\centering
\includegraphics[width=0.9\linewidth]{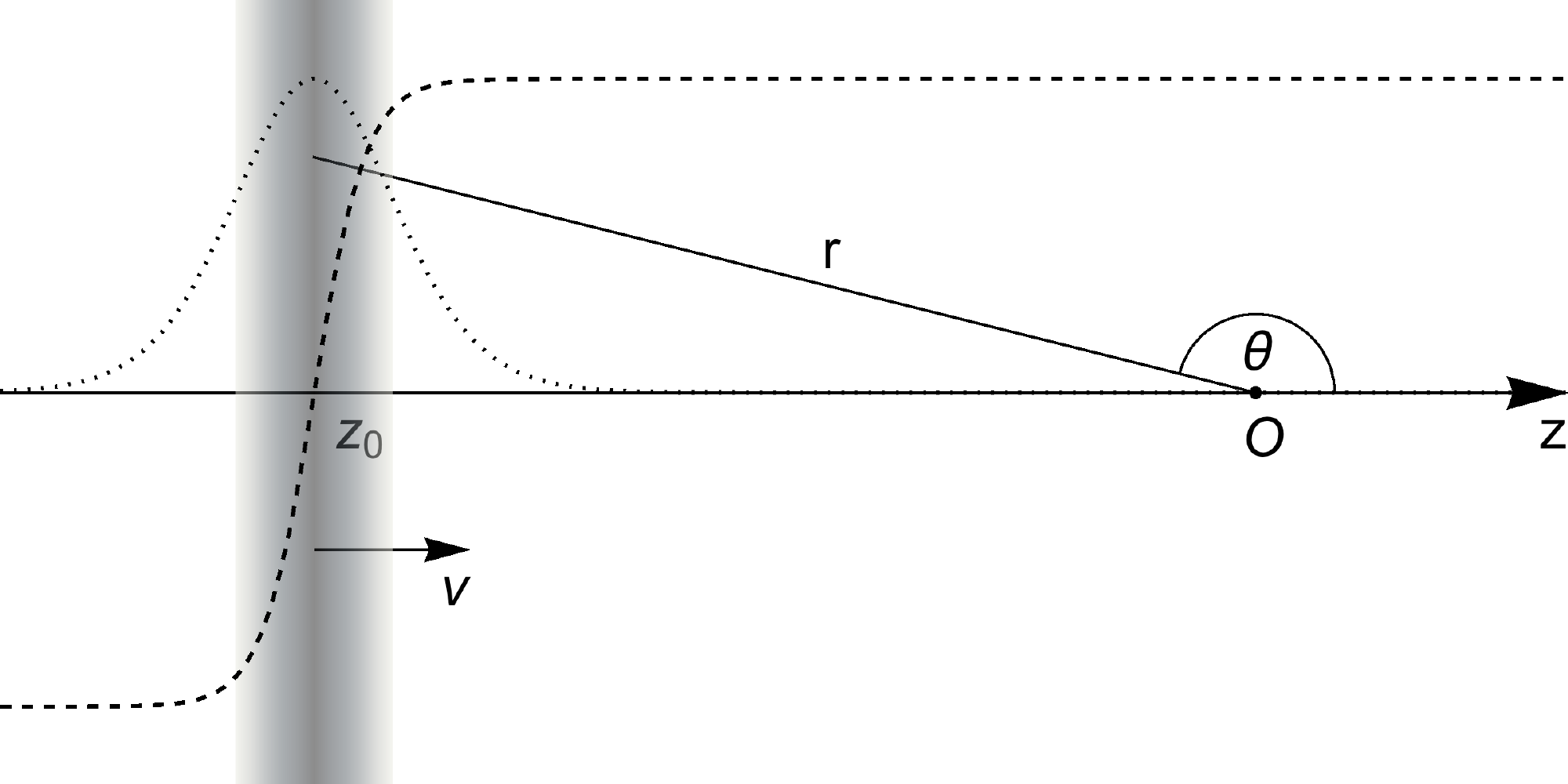}
\caption{\label{fig:spherical}A planar domain wall in spherical coordinates. The origin of the coordinate system is denoted as $O$; $z_0$ is the initial location of the domain wall with respect to the origin. The dashed and dotted plots show the value of the field and the energy density of the field along the $z$ axis, respectively.}
\end{figure}

In the following we will assume $d = 3$ and work in spherical coordinates $(r,\theta,\varphi)$. Accordingly, it is instructive to express Eq.\ (\ref{eqn:SolMinkd}) in spherical coordinates; this expression will also serve as initial data in our simulations. Choosing the axis of the system $(r,\theta,\varphi)$ to be aligned with the $z$ axis, we get
\begin{align} 
\label{eqn:3dMink}
\phi(t,r,\theta,\varphi)&=\eta\tanh\left(\eta\sqrt{\frac{\lambda}{2}}\frac{(r \cos \theta- z_0-v t)}{\sqrt{1-v^2}}\right).
\end{align}
The time derivative of the field is given by
\begin{eqnarray} 
\label{eqn:3dMinkPrim}
\lefteqn{\partial_t \phi(t,r,\theta,\varphi)=} \nonumber \\
&& -\sqrt{\frac{\lambda}{2}} \frac{1}{\sqrt{1-v^2}} \frac{\eta^2 v}{\cosh^2\left(\eta\sqrt{\frac{\lambda}{2}}\frac{r\cos\theta-z_0-v t}{\sqrt{1-v^2}}\right)}.
\end{eqnarray}
The corresponding geometry is illustrated in Fig.\ \ref{fig:spherical}.

\section{Scalar fields in the Kerr spacetime}

There is no unique way to postulate the equation of motion for the scalar field in a curved spacetime. In this paper we assume the minimal coupling. Since we only deal with spacetimes characterized by the vanishing scalar curvature (vacuum or electro-vacuum), the assumed minimal coupling also coincides with another common choice --- the conformal coupling \cite{Wal, Son93}.

The Lagrangian density of the minimally coupled scalar field has the form
\begin{equation} 
\mathcal{L}=-\frac{1}{2}\nabla_\mu \phi \nabla^\mu \phi-V(\phi),
\label{eqn:LagCur}
\end{equation}
where $\nabla_\mu$ denotes the covariant derivative. The corresponding equation of motion reads
\begin{equation} 
\nabla_{\mu}\nabla^{\mu}\phi-V'(\phi)=0,
\label{eqn:EomCur}
\end{equation}
and the energy-momentum tensor of the scalar field can be written as \cite{Wal}
\begin{equation} 
T_{\mu\nu}=\nabla_\mu \phi \nabla_\nu \phi-g_{\mu\nu}\left(\frac{1}{2}\nabla_\sigma \phi \nabla^\sigma\phi+V(\phi)\right).
\label{eqn:TmunuCur}
\end{equation}

The above energy-momentum tensor satisfies the dominant energy condition. For any future directed timelike vector $X^\nu$, the vector field $Y^\mu=-{T^\mu}_\nu X^\nu$ is future directed and timelike or null. To show this, it suffices to choose a locally inertial frame and perform appropriate Lorentz transformations, so that (locally) $X^\nu=(1,0,0,0)$. Then
\begin{align} 
Y^\mu=-{T^\mu}_0=&-\partial^\mu\phi \partial_0\phi+\delta^\mu_0\left(-\frac{1}{2}\left(\partial_0 \phi\right)^2\right.\nonumber\\
&\left.+\frac{1}{2} \partial_i \phi \; \partial^i \phi+V(\phi)\right).
\label{eqn:DEC1}
\end{align}
We see that $Y^0 \geq 0$ for non-negative potentials $V(\phi)$.  Moreover, rewriting the Lagrangian density given by Eq.\ (\ref{eqn:LagCur}) in a locally inertial frame, we obtain
\begin{align} 
Y^\mu Y_\mu&=(\partial_0\phi)^2\partial_\mu\phi\partial^\mu\phi+2(\partial_0\phi)^2 \mathcal{L}-\mathcal{L}^2\nonumber\\
&=-2(\partial_0\phi)^2 V(\phi)-\mathcal{L}^2\leq 0.
\label{eqn:DEC2}
\end{align}
The dominant energy condition assures that the energy flow of matter is always slower than the speed of light \cite{Wal}.

In this paper we consider the Kerr metric written in the Kerr-Schild coordinates $(t,r,\theta,\varphi)$. The line element reads
\begin{align} 
ds^2=&-\left(1-\frac{2m r}{\Sigma}\right)dt^2+\frac{4m r}{\Sigma}dt dr-\frac{4a m r \sin^2 \theta}{\Sigma}dt d\varphi\nonumber \\
&+\left(1+\frac{2m r}{\Sigma}\right)dr^2-2a\sin^2 \theta \left(1+\frac{2m r}{\Sigma}\right)dr d\varphi\nonumber \\
&+\Sigma d\theta^2+\sin^2\theta \left(\Sigma+a^2\left(1+\frac{2m r}{\Sigma}\right) \sin^2\theta\right)d\varphi^2,
\label{eqn:Kerr}
\end{align}
where
\begin{align}
\Sigma&=r^2+a^2\cos^2 \theta, \label{eqn:Sigma}
\end{align}
and the ranges of the variables are given by
\begin{align}
t\in(-\infty,+\infty),\quad r\in(0,\infty),\quad \theta\in(0,\pi),\quad \varphi\in(0,2\pi).
\label{eqn:ranges}
\end{align}
The parameters $m$ and $a$ are interpreted as the mass of the black hole and the angular momentum (spin) parameter, respectively. The advantage of using the Kerr-Schild coordinates is that they are regular at the horizons ($r_\pm=m\pm\sqrt{m^2-a^2}$). The physical singularity at $r=0$, $\theta=\pi/2$ is still present \cite{Wal}.

The energy density of the scalar field is defined as $\rho=T_{\mu\nu}n^\mu n^\nu$, where $n^\mu$ is a normalized time-like vector orthogonal to the hypersurfaces of constant time. We get, for metric (\ref{eqn:Kerr}),
\begin{align}
n^\mu=\left(\sqrt{1+\frac{2m r}{\Sigma}},-\frac{2mr}{\Sigma\sqrt{1+\frac{2mr}{\Sigma}}},0,0\right),
\label{eqn:vector}
\end{align}
and
\begin{align}
\rho=&\frac{1}{\Sigma}\left[\frac{1}{2}(\Sigma+2mr)(\partial_t \phi)^2-2mr\partial_t\phi\partial_r\phi\right.\nonumber\\
&\left.+\frac{1}{2}\left(a^2\sin^2\theta+\frac{\Sigma^2+4m^2 r^2}{\Sigma+2mr}\right)(\partial_r\phi)^2+\frac{1}{2}(\partial_\theta\phi)^2\right.\nonumber\\
&\left.+a\partial_r\phi\partial_\varphi\phi+\frac{1}{2\sin^2\theta}(\partial_\varphi\phi)^2\right]+V(\phi).
\label{eqn:edensity}
\end{align}
Of course $\rho\geq0$, since the energy-momentum tensor satisfies the dominant energy condition.

The d'Alembert operator written in Kerr-Schild coordinates reads
\begin{align} 
\Box\phi=\frac{1}{\Sigma}&\left[-(\Sigma+2mr) \partial^2_t \phi+2m\partial_t \phi +4m r\partial_t\partial_r \phi\right. \nonumber \\ 
&\left. +\partial_r( \Delta \partial_r\phi)+\frac{1}{\sin\theta}\partial_\theta(\sin\theta\partial_\theta\phi)+2a \partial_r\partial_\varphi \phi \right.\nonumber\\
&\left.+\frac{1}{\sin^2\theta}\partial_\varphi\partial_\varphi\phi\right],
\label{eqn:DeLambert}
\end{align}
where $\Box=\nabla_\mu \nabla^\mu$ and
\begin{align}
\Delta&=r^2-2m r+a^2. \label{eqn:Delta}
\end{align}

In the following we assume the axial symmetry. Equations (\ref{eqn:DeLambert}), (\ref{eqn:EomCur}), and (\ref{eqn:Potential4}) yield the equation of motion
\begin{align} 
\partial^2_t \phi=&\frac{1}{\Sigma+2mr}\left[2m\partial_t \phi +4m r\partial_t\partial_r \phi+\partial_r( \Delta \partial_r\phi)\right.\nonumber\\
&\left.+\frac{1}{\sin\theta}\partial_\theta(\sin\theta\partial_\theta\phi)
+\Sigma\lambda\eta^2 \phi-\Sigma\lambda\phi^3\right].
\label{eqn:EomFin}
\end{align}

The equations of characteristics of Eq.\ (\ref{eqn:EomFin}) in the radial direction can be found as
\begin{align} 
&r'(t)=\frac{1}{r^2+2m r+a^2 \cos^2 \theta}\left(-2m r\right.\nonumber\\
&\left.\pm \sqrt{r^4+a^4\cos^2 \theta+a^2 r^2 (1+\cos^2\theta)+2m a^2 r \sin^2\theta}\right).
\label{eqn:charr}
\end{align}
The corresponding equation in the angular direction can be written as
\begin{equation} 
\theta'(t)=\pm\frac{1}{\sqrt{r^2+2m r+a^2 \cos^2 \theta}}.
\label{eqn:chartheta}
\end{equation}
One of expressions (\ref{eqn:charr}) is always negative (it refers to the ingoing characteristic); the other one (referring to the outgoing characteristic) is positive for $r > r_+$ or $r < r_-$ and negative for $r_- < r <r_+$. It is a direct consequence of the causal structure of Eq.\ (\ref{eqn:EomFin}). We will use this fact in the numerical implementation of our simulations. Analogous expressions corresponding to the angular characteristics (\ref{eqn:chartheta}) have positive and negative values.

\section{Numerical scheme}

\begin{figure}
\centering
\includegraphics[width=0.9\linewidth]{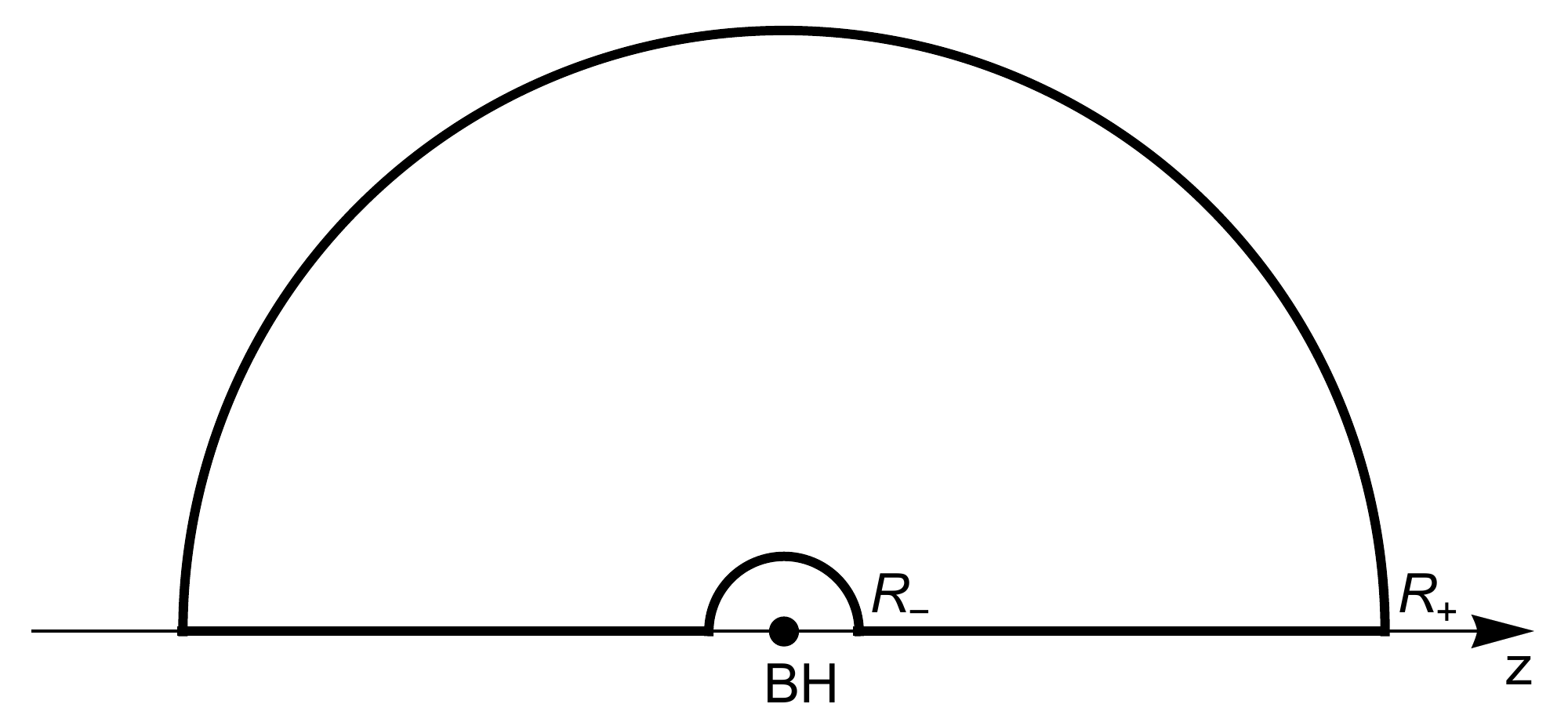}
\caption{\label{fig:domain}The shape of the numerical domain.}
\end{figure}

We solve Eq. (\ref{eqn:EomFin}) using a variant of the so-called method of lines. Our numerical grid spans the region $R_- \le r \le R_+$, $0 \le \theta \le \pi$, where $R_-$ and $R_+$ denote the locations of the inner and outer edges of the grid, respectively. This is illustrated in Fig.\ \ref{fig:domain}. The choice of $R_+$ is arbitrary as long as it is much larger than the radius of the horizon. For the inner boundary we choose $R_- = m$, so that $r_+ > R_- > r_-$. As a result, we do not need to impose any boundary conditions at $r = R_-$. We will return to this point later in this section. The parameters of grids used in the simulations are presented in Table \ref{tab:gridparameters}. Set I was our default set of parameters used in most simulations, while Set II was used to produce the plots shown in this paper. Our grids are equidistant in both directions $r$ and $\theta$. We also set $m = 1$ in all our simulations.

\begin{table}
\caption{\label{tab:gridparameters}Parameters of grids used in the simulations. Set I was the default set of parameters. Set II was used to prepare the plots included in this paper.}
\begin{ruledtabular}
\begin{tabular}{ccc}
&Set I& Set II\\ \hline
inner boundary $R_-$ & 1.0 & 1.0 \\
outer boundary $R_+$ & 20.0 & 40.0 \\
number of radial zones & 600 & 1200 \\
number of axial zones & 400 & 800
\end{tabular}
\end{ruledtabular}
\end{table}

\begin{figure}
\centering
\includegraphics[width=0.9\linewidth]{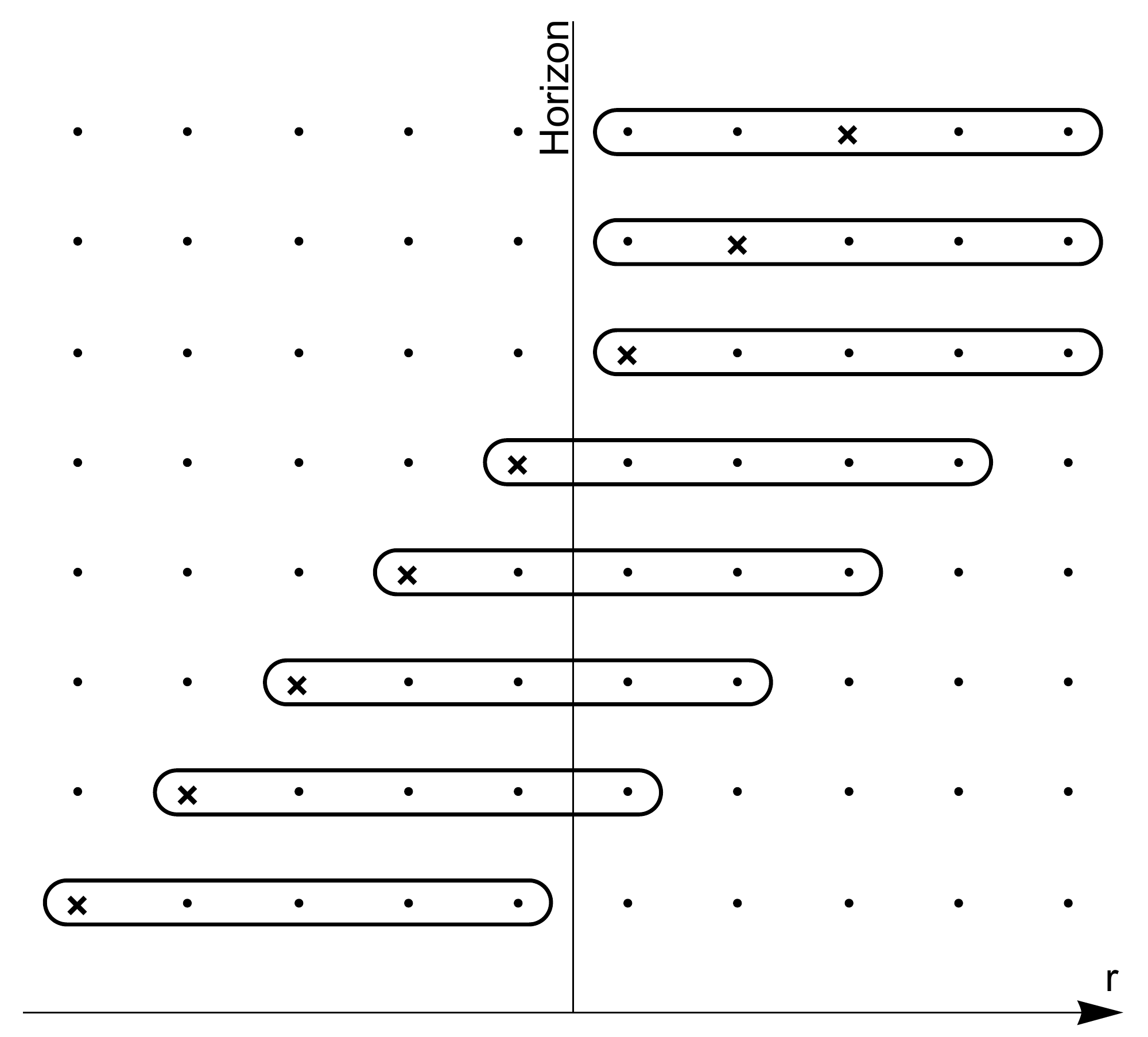}\\
\caption{\label{fig:5discret}An illustration of the radial discretisation scheme used to ensure the proper causal behavior of solutions. The rows depict appropriate stencils used to compute numerical derivatives at points denoted by a cross. They correspond to Eqs.\ (\ref{eqn:5point-twoside}--\ref{eqn:5point-oneside}) in the text.} 
\end{figure}

Equation (\ref{eqn:EomFin}) is first transformed into a system of first-order-in-time partial differential equations
\begin{subequations}
\label{eqn:EomCN}
\begin{align}
\partial_t u_{(1)}=&u_{(2)}, \label{eqn:EomCNa} \\
\partial_t u_{(2)}=& \frac{1}{\Sigma+2mr}\left[2m u_{(2)} +4m r \partial_r u_{(2)}+\partial_r( \Delta \partial_r u_{(1)})\right.\nonumber\\
&\left. +\frac{1}{\sin\theta}\partial_\theta(\sin\theta\partial_\theta u_{(1)})+\Sigma\lambda\eta^2 u_{(1)}-\Sigma\lambda u_{(1)}^3\right], \label{eqn:EomCNb}
\end{align}
\end{subequations}
where $u_{(1)}:=\phi$, $u_{(2)}:=\partial_t \phi$. These equations are discretised in the spatial directions $(r,\theta)$, taking into account their causal structure. We use 5-point finite difference formulas. The discretisation in the radial direction requires an explanation. Let $u_{i,j} = u(r_i,\theta_j)$ denote the value of a function $u$ in the node of the grid labelled by indices $(i,j)$. For $r_i \gg r_+$ we use standard, centered 5-point formulas
\begin{subequations}
\label{eqn:5point-twoside}
\begin{align}
\partial_r u_{i,j} =&\frac{-u_{i+2,j}+8u_{i+1,j}-8u_{i-1,j}+u_{i-2,j}}{12\Delta r}+\mathcal{O}(\Delta r^4),\\
\partial_r^2 u_{i,j} =&\frac{-u_{i+2,j}+16u_{i+1,j}-30u_{i,j}+16u_{i-1,j}-u_{i+2,j}}{12\Delta r^2}\nonumber\\
&+\mathcal{O}(\Delta r^4),
\end{align}
\end{subequations}
where $\Delta r$ is the radial distance between two grid nodes. Near and below the horizon we use the following asymmetric 5-point formulas:
\begin{subequations}
\label{eqn:5point-oneandhalfside}
\begin{align}
\partial_r u_{i,j} =&\frac{u_{i+3,j}-6u_{i+2,j}+18u_{i+1,j}-10u_{i,j}-3u_{i-1,j}}{12\Delta r} \nonumber \\
& +\mathcal{O}( \Delta r^4 ),\\
\partial_r^2 u_{i,j} =&\frac{-u_{i+3,j}+4u_{i+2,j}+6u_{i+1,j}-20u_{i,j}+11u_{i-1,j}}{12\Delta r^2}\nonumber\\
&+\mathcal{O}(\Delta r^3),
\end{align}
\end{subequations}
\begin{subequations}
\label{eqn:5point-oneside}
\begin{eqnarray}
\lefteqn{ \partial_r u_{i,j} = } \nonumber \\
&& \frac{-3u_{i+4,j}+16u_{i+3,j}-36u_{i+2,j}+48u_{i+1,j}-25u_{i,j}}{12\Delta r}\nonumber\\
&& +\mathcal{O}(\Delta r^4),\\
\lefteqn{\partial_r^2 u_{i,j} = } \nonumber \\
&& \frac{11u_{i+4,j}-56u_{i+3,j}+114u_{i+2,j}-104u_{i+1,j}+35u_{i,j}}{12\Delta r^2}\nonumber\\
&& +\mathcal{O}(\Delta r^3).
\end{eqnarray}
\end{subequations}
Figure \ref{fig:5discret} demonstrates the way in which these formulas are used. In practice, no value of the field from the black hole region influences the values outside the horizon. Also, no boundary condition is necessary at $r = R_-$. The discretisation in the angular direction is straightforward; we use centered finite difference formulas.

Discretised Eqs.\ (\ref{eqn:EomCN}) are solved using the Crank-Nicolson method \cite{Tho, Pre}. We use this method iteratively \cite{Teu00}, as described in Appendix \ref{sec:cn}.

For the initial data we choose a planar domain wall perpendicular to the $z$ axis (the symmetry axis of the spacetime), located at a large distance from the black hole, and moving with the velocity $v$ (see Fig.\ \ref{fig:spherical}). Since the Kerr metric is asymptotically flat, we expect that at large distances the solution describing a planar domain wall should resemble the analytic solution characteristic for the Minkowski spacetime, given by Eqs.\ (\ref{eqn:3dMink}) and (\ref{eqn:3dMinkPrim}). Consequently, we assume as initial data the expressions:
\begin{subequations}
\label{eqn:initial1}
\begin{align} 
\phi(t,r,\theta)|_{t=0}=&\eta\tanh\left(\eta\sqrt{\frac{\lambda}{2}}\frac{(r \cos \theta- z_0)}{\sqrt{1-v^2}}\right),\\
\partial_t \phi(t,r,\theta)|_{t=0}=&-\eta^2 v \sqrt{\frac{\lambda}{2}} \frac{1}{\sqrt{1-v^2}} \nonumber\\
&\times\frac{1}{\cosh^2\left(\eta\sqrt{\frac{\lambda}{2}}\frac{r \cos\theta-z_0}{\sqrt{1-v^2}}\right)}.
\end{align}
\end{subequations}

Analytic expressions (\ref{eqn:3dMink}) and (\ref{eqn:3dMinkPrim})  are also used to provide the outer boundary conditions at $r = R_+$. In explicit terms, we set
\begin{align} 
\phi(t,r,\theta)|_{r=R_+}=\eta\tanh\left(\eta\sqrt{\frac{\lambda}{2}}\frac{(R_+ \cos \theta- z_0-v t)}{\sqrt{1-v^2}}\right).
\label{eqn:bc1}
\end{align}
At the $z$ axis, which is also a boundary of the numerical domain, we impose standard regularity (von Neumann) conditions:
\begin{align} 
\partial_\theta\phi(t,r,\theta)|_{\theta=0}=\partial_\theta\phi(t,r,\theta)|_{\theta=\pi}=0.
\label{eqn:bc2}
\end{align}

We tested our numerical scheme by simulating the evolution of a domain wall in the Minkowski spacetime. In this case we imposed the inner boundary conditions (at $r=R_-$) in the exact form
\begin{align} 
\phi(t,r,\theta)|_{r=R_-}=\eta\tanh\left(\eta\sqrt{\frac{\lambda}{2}}\frac{(R_- \cos \theta- z_0-v t)}{\sqrt{1-v^2}}\right).
\label{eqn:bc3}
\end{align}
The obtained numerical solutions agree with the exact formula (\ref{eqn:3dMink}).

\section{Results}\label{sec:results}

Without the loss of generality, we assume $m = 1$, i.e., all distances are expressed in the units of $m$ (of course, this also affects the system of units used for the field variables).

We set $\eta=0.1$ and $\lambda=100$ in all our simulations. It is shown in Appendix \ref{sec:paren} that this choice ensures that the local energy of the domain wall is much less than the energy associated with the black hole.

Our simulations are limited to velocities $v > 1/2$. There are two reasons for that. The first one is that domain walls seem to be repulsed by the black hole, and this effect is more pronounced for smaller initial velocities $v$. We show in Appendix \ref{sec:rep} that a domain wall that is initially at rest drifts away from the black hole. This repulsion seems to be a physical effect rather than a property of the used coordinate system. For simplicity, we use standard Boyer-Lindquist coordinates in Appendix \ref{sec:rep}. The second reason is that smaller values of $v$ would require much larger numerical domains, in order to minimize the spurious influence of the outer boundary conditions. Parameters $v$ and $a$ (the spin parameter of the black hole) used in our simulations are collected in Table \ref{tab:parameters}.

\begin{table*}
\caption{\label{tab:parameters}Parameters of the simulations.}
\begin{ruledtabular}
\begin{tabular}{cccccccccccc}
Initial velocity $v$ & 0.5 & 0.6 & 0.7 & 0.8 & 0.9 & 0.93 & 0.95 & 0.97 & 0.99 & 0.995 & 0.999\\
Black hole spin parameter $a$ & 0 & 0 & 0 & 0 & 0 & 0 & 0 & 0 & 0 & 0 & 0 \\ \hline
Initial velocity $v$ & 0.9 & 0.9 & 0.9 & 0.9 & 0.9 & 0.9 & 0.9 & 0.9 & 0.9 & 0.9 & 0.9\\
Black hole spin parameter $a$ & 0.5 & 0.6 & 0.7 & 0.8 & 0.9 & 0.93 & 0.95 & 0.97 & 0.99 & 0.995 & 0.999\\
\end{tabular}
\end{ruledtabular}
\end{table*}

\begin{figure*}
\subfloat[\label{fig:s1}]{\includegraphics[width=.4\linewidth]{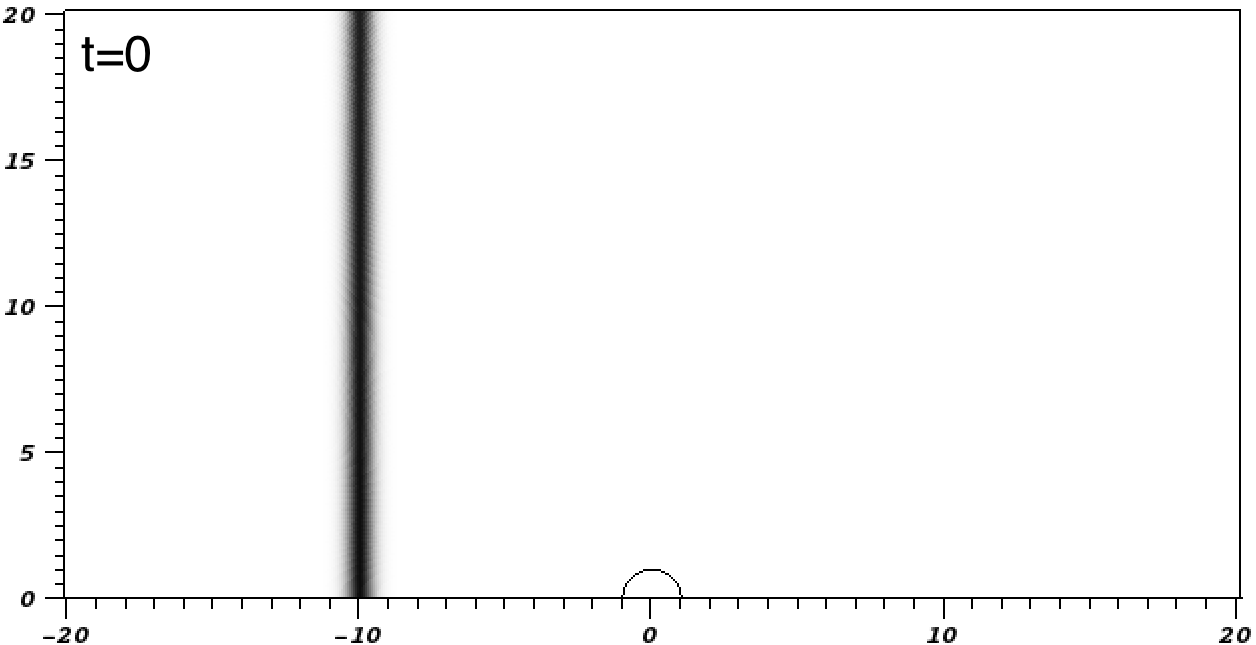}}
\subfloat[\label{fig:s2}]{\includegraphics[width=.4\linewidth]{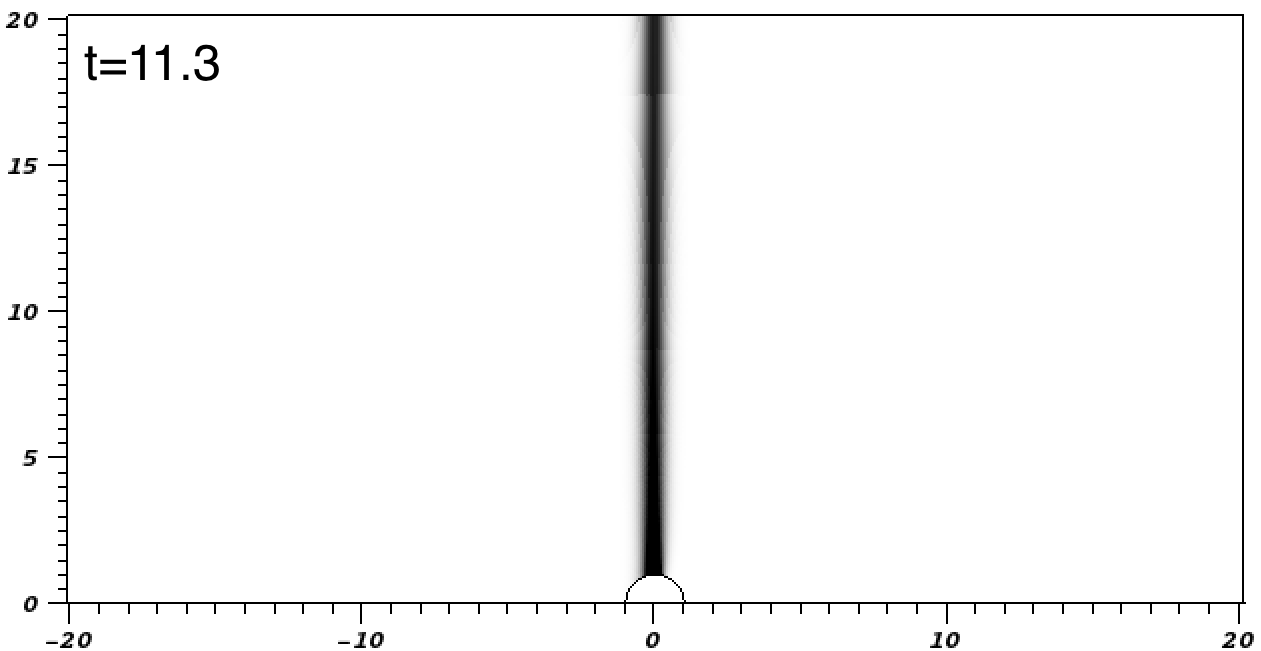}}\hfill
\subfloat[\label{fig:s3}]{\includegraphics[width=.4\linewidth]{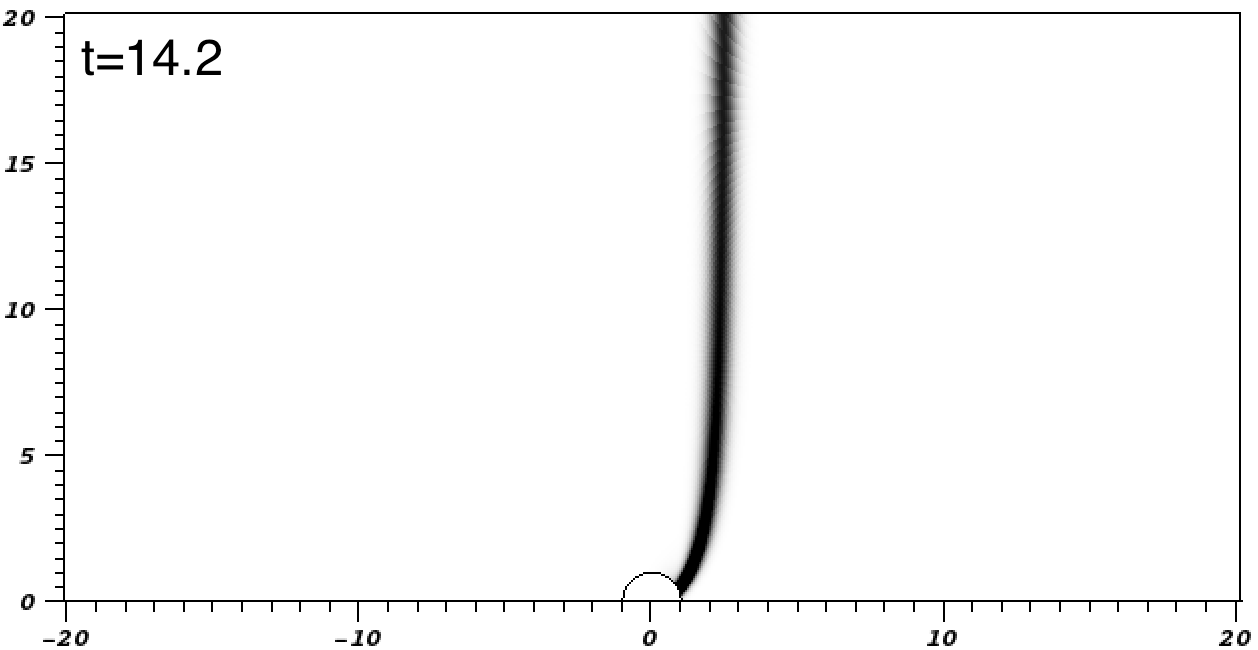}}
\subfloat[\label{fig:s4}]{\includegraphics[width=.4\linewidth]{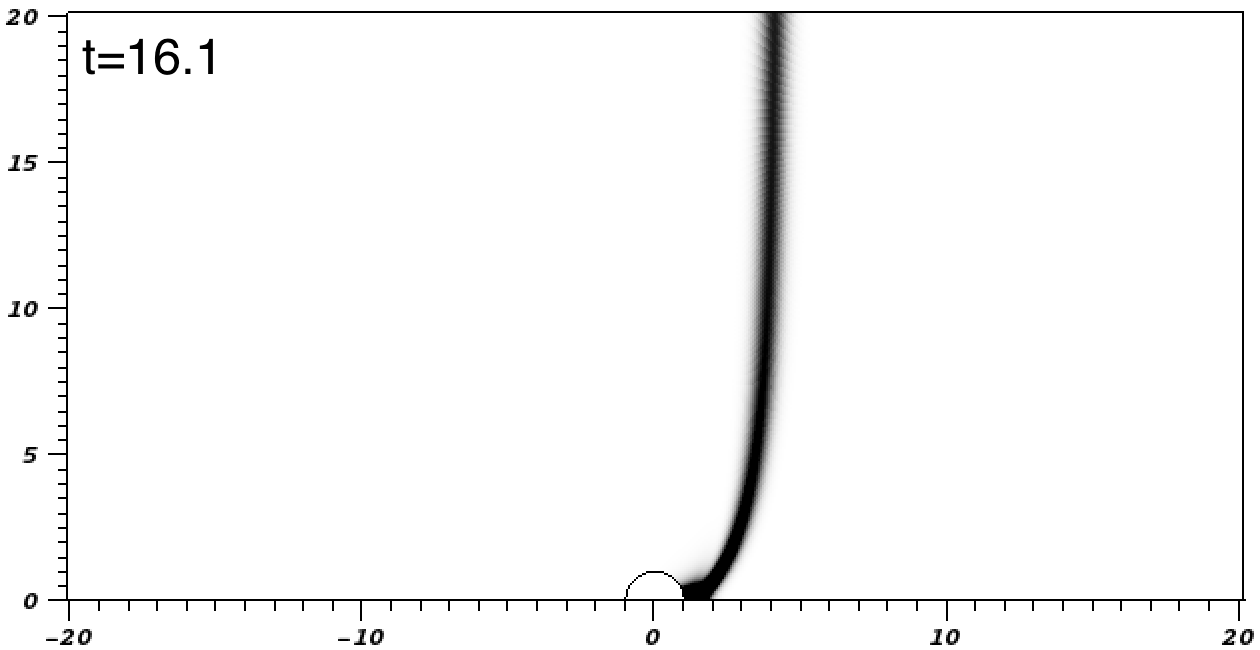}}\hfill
\subfloat[\label{fig:s5}]{\includegraphics[width=.4\linewidth]{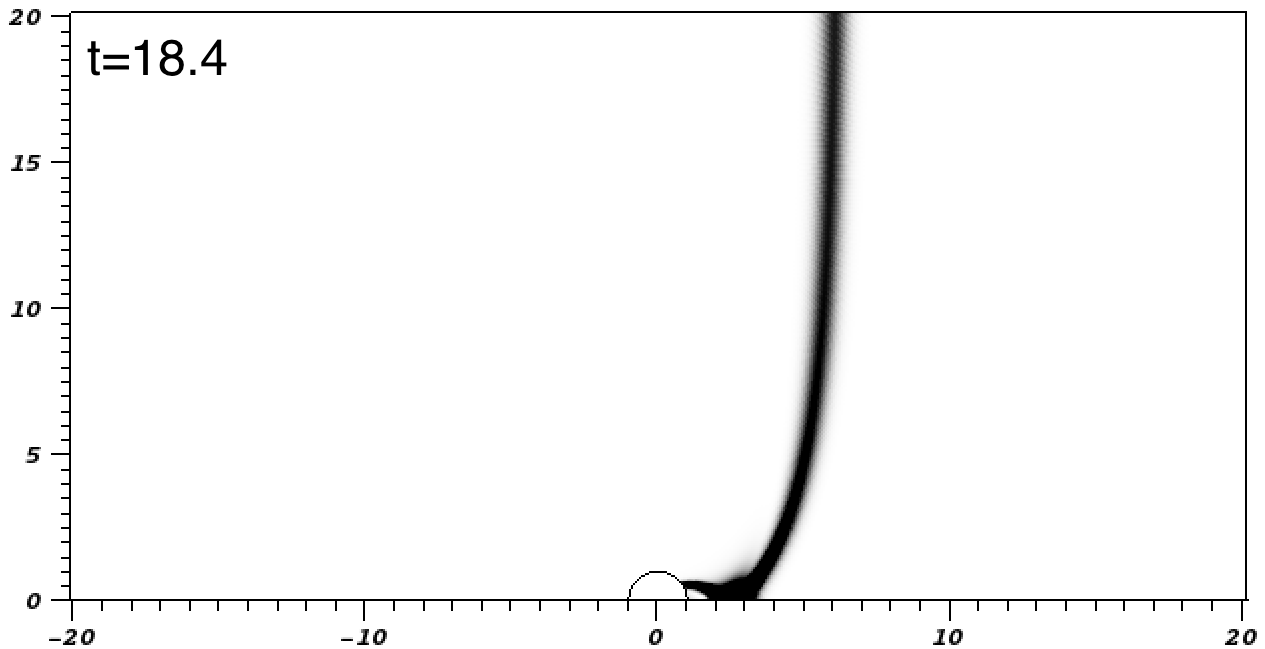}}
\subfloat[\label{fig:s6}]{\includegraphics[width=.4\linewidth]{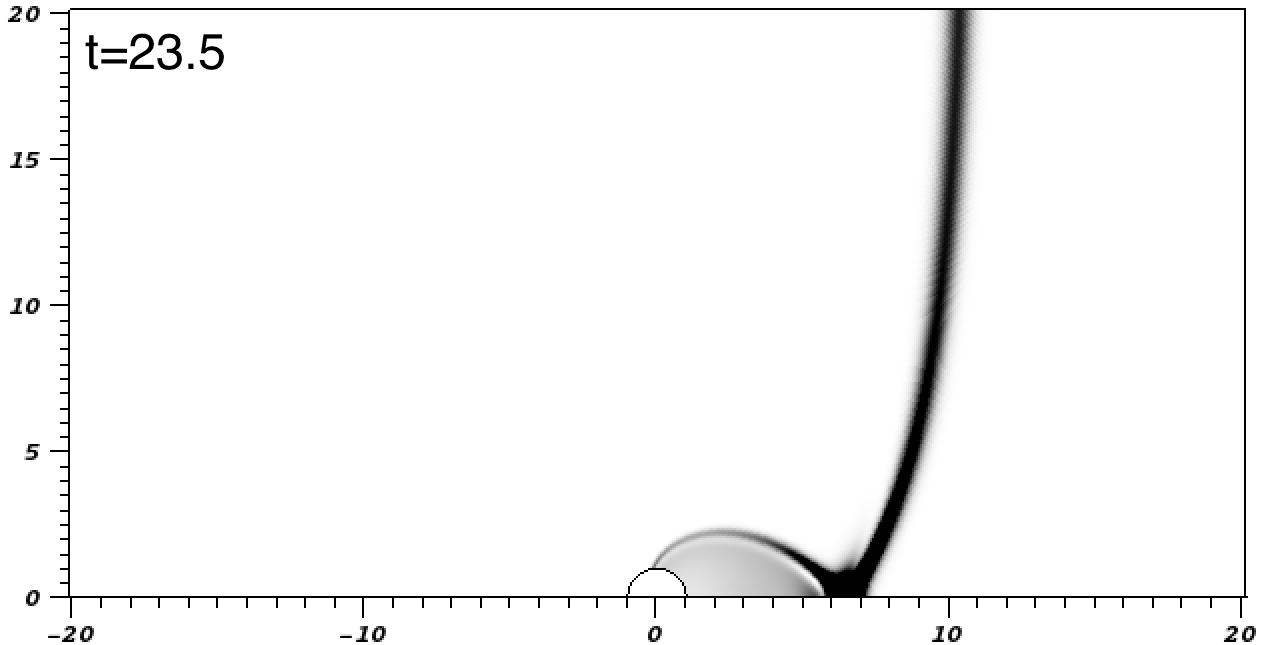}}\hfill
\subfloat[\label{fig:s7}]{\includegraphics[width=.4\linewidth]{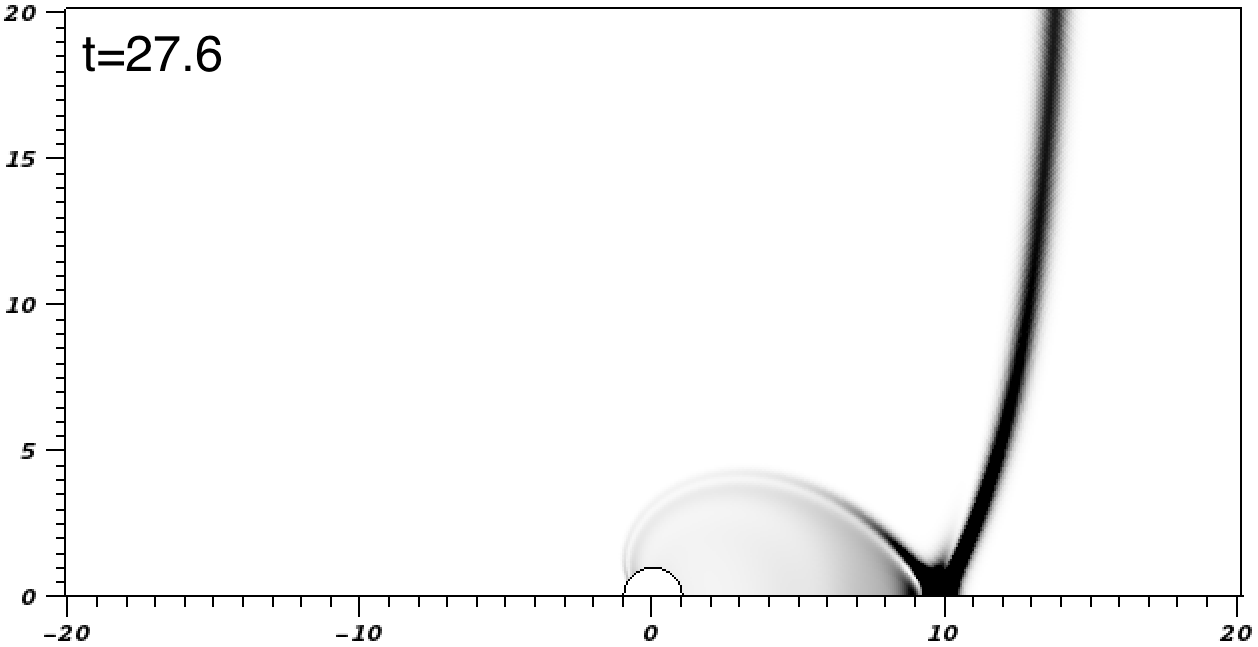}}
\subfloat[\label{fig:s8}]{\includegraphics[width=.4\linewidth]{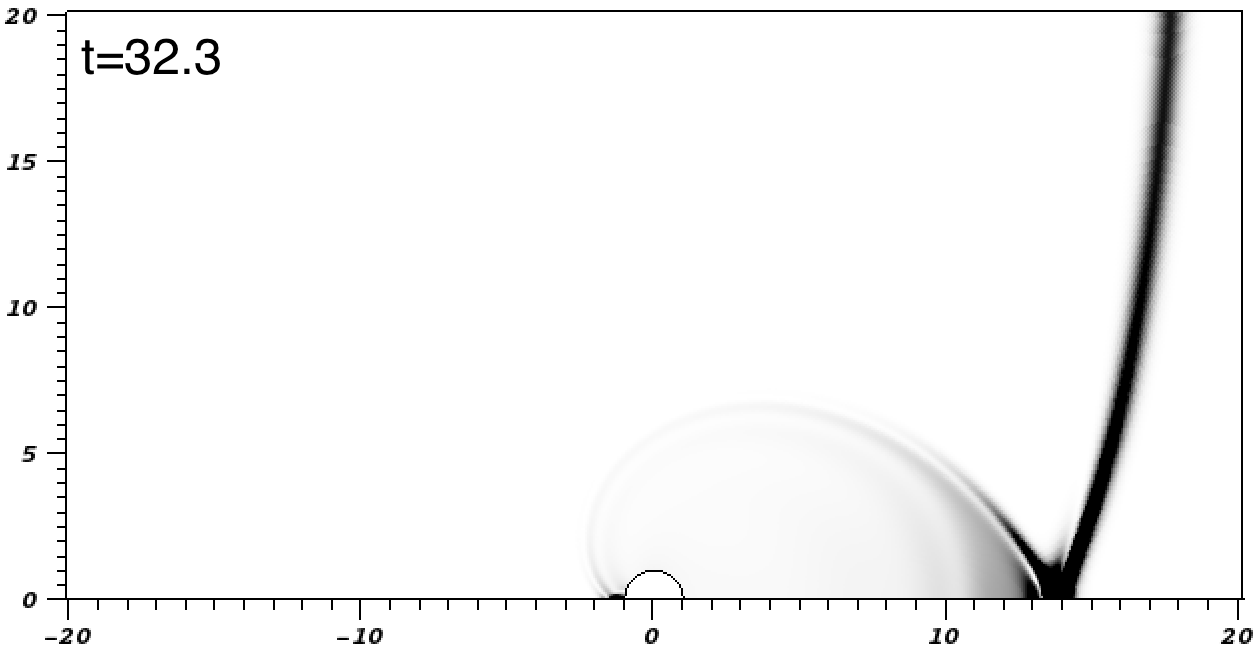}}\hfill
\subfloat[\label{fig:s9}]{\includegraphics[width=.4\linewidth]{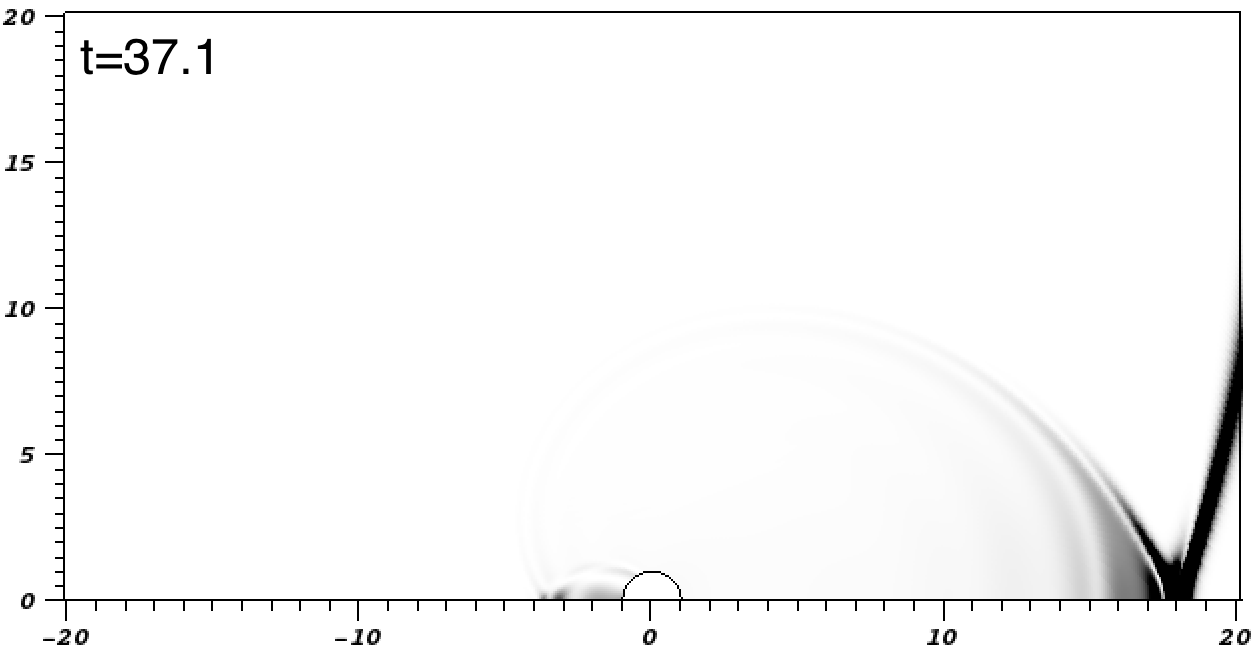}}
\subfloat[\label{fig:s10}]{\includegraphics[width=.4\linewidth]{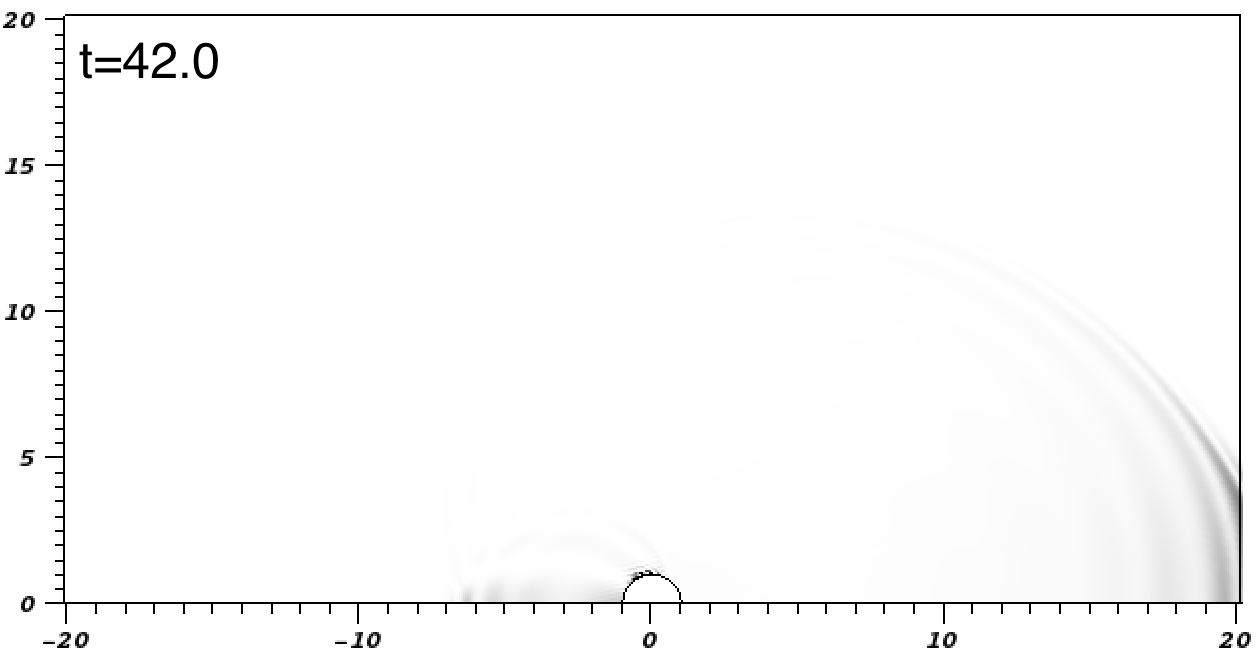}}\hfill
\caption{Successive steps of the evolution of the domain wall with the initial velocity $v=0.9$ near the Schwarzschild black hole of mass $m=1$. Dark shades denote regions with higher energy density (\ref{eqn:edensity}).}
\label{fig:schwarzschild}
\end{figure*}

\begin{figure*}
\subfloat[\label{fig:stab1}]{\includegraphics[width=.45\linewidth]{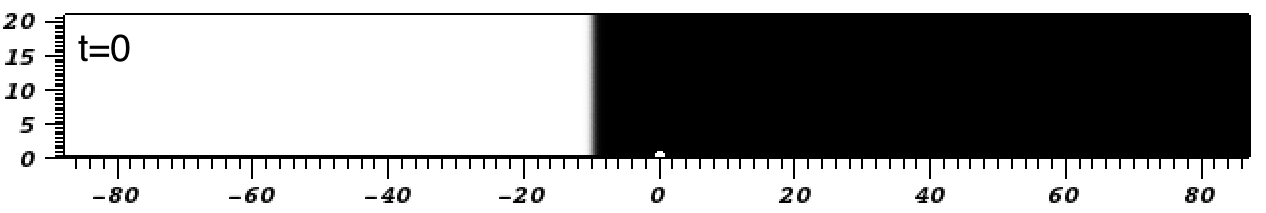}}
\subfloat[\label{fig:stab2}]{\includegraphics[width=.45\linewidth]{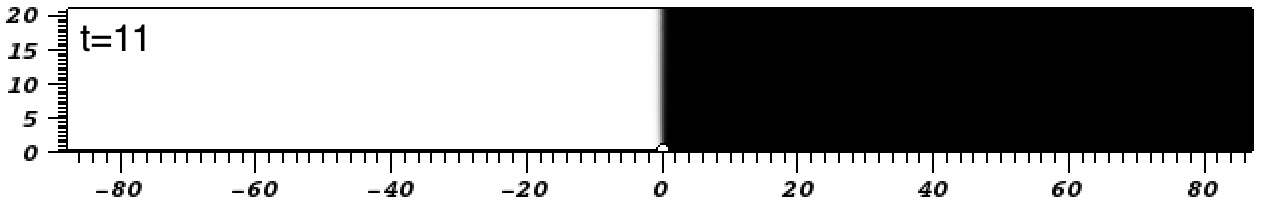}}\hfill
\subfloat[\label{fig:stab3}]{\includegraphics[width=.45\linewidth]{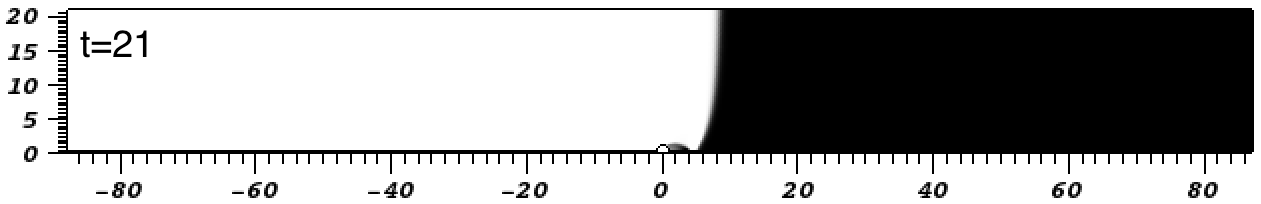}}
\subfloat[\label{fig:stab4}]{\includegraphics[width=.45\linewidth]{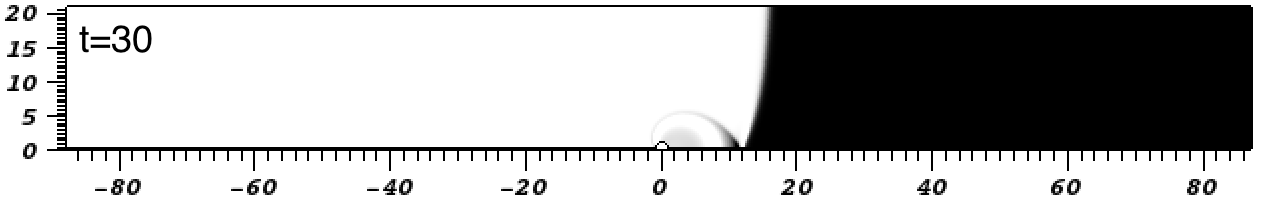}}\hfill
\subfloat[\label{fig:stab5}]{\includegraphics[width=.45\linewidth]{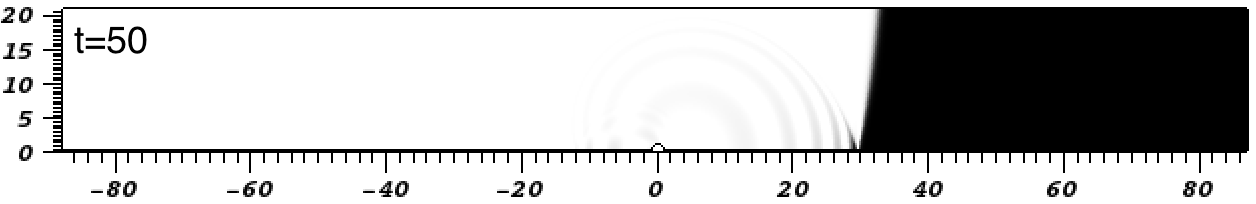}}
\subfloat[\label{fig:stab6}]{\includegraphics[width=.45\linewidth]{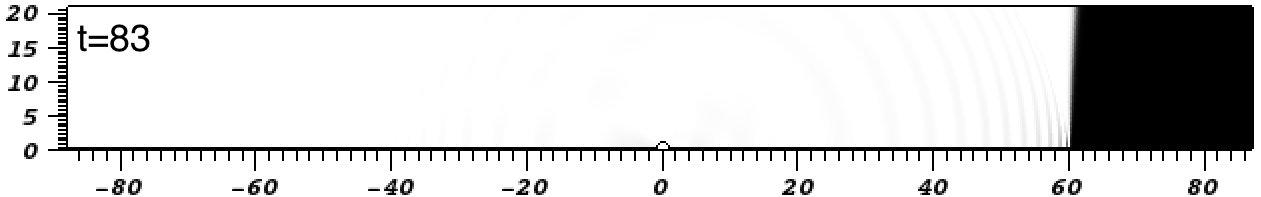}}\hfill
\caption{Changes in the shape of the domain wall with the initial velocity $v=0.9$ during the transit through the Schwarzschild black hole of mass $m=1$. White and black regions refer to the field in one of two domains.}
\label{fig:stab}
\end{figure*}

An example of the evolution of a domain wall on the Schwarzschild background is shown in Fig.\ \ref{fig:schwarzschild}, where we plot the field energy density. Figure \ref{fig:s1} shows the initial data --- a domain wall with the initial velocity $v=0.9$ located at the distance  $z_0 = 10$ from the black hole. The domain wall moves almost undisturbed (Fig.\ \ref{fig:s2}) until it passes through the black hole (Fig.\ \ref{fig:s3}). This results in an excitation of the field (Fig.\ \ref{fig:s4}) followed by a creation of a separate domain (Fig.\ \ref{fig:s5}) and another domain wall, which finally encompasses the black hole (Figs.\ \ref{fig:s6}, \ref{fig:s7}, \ref{fig:s8}). This new domain wall is reflected when it reaches the opposite axis of symmetry; we observe a kind of ringing (Fig.\ \ref{fig:s9},\ref{fig:s10}). The similar structure has been observed in higher-dimensional brane-dilaton-black-hole systems \cite{Nak16}. The long time structure of these ringings can be observed in Fig.\ \ref{fig:stab}.

\begin{figure*}
\subfloat[\label{fig:v1}]{\includegraphics[width=.4\linewidth]{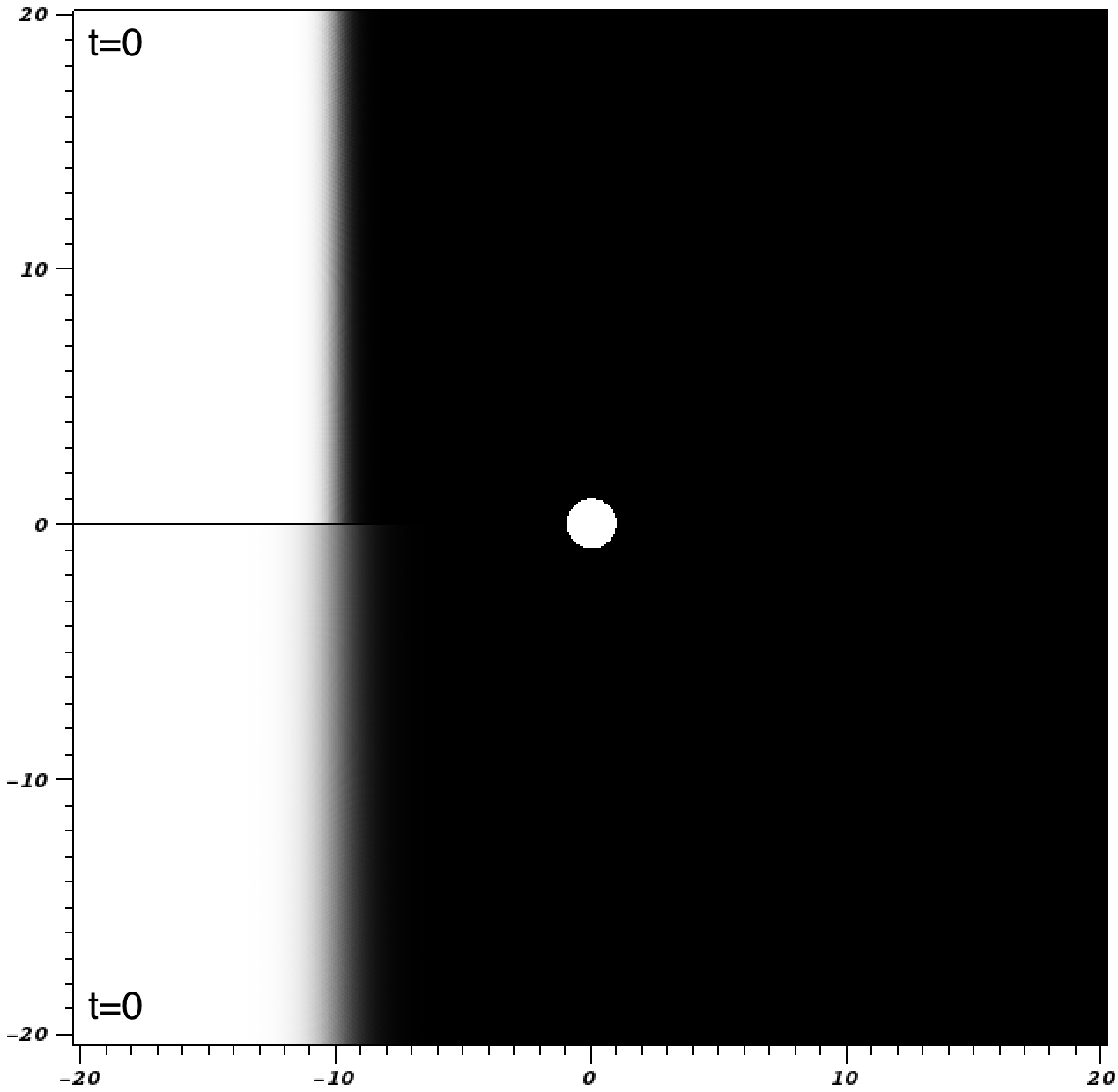}}
\subfloat[\label{fig:v2}]{\includegraphics[width=.4\linewidth]{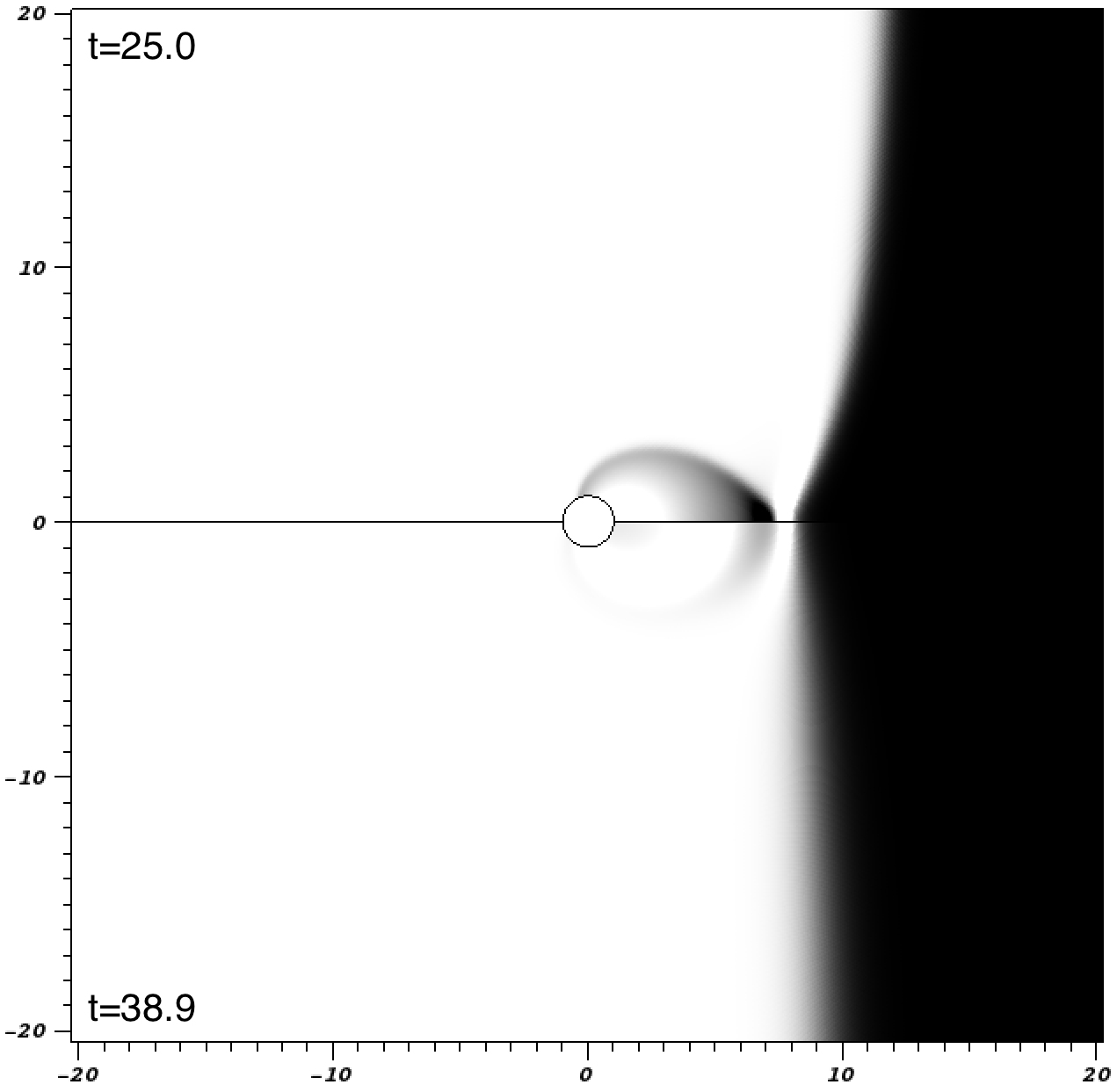}}\hfill
\caption{Successive steps of the evolution of domain walls with the initial velocities $v=0.9$ (upper halves) and $v=0.5$ (lower halves) near the Schwarzschild black hole of mass $m=1$. White and black regions refer to the field in one of two domains.}
\label{fig:velocity}
\end{figure*}

Simulations of the domain wall evolution near the Schwarzschild black hole have been repeated for a range of initial velocities presented in Table \ref{tab:parameters}. The comparison of exemplary results is presented in Fig.\ \ref{fig:velocity}. The initial setup is shown in Fig.\ \ref{fig:v1} --- we observe that faster domain walls appear more compact due to the Lorentz contraction. In Figure \ref{fig:v2} we present two domain walls at distinct instances of time referring to a similar state of their evolution. The additional ringing structure and the distortion of the domain wall caused by the transition through the black hole are much weaker in the case with the lower initial velocity $v$. This observation is also confirmed by the investigation of other cases with the parameters collected in Table \ref{tab:parameters} --- domain walls moving initially with larger velocities are perturbed stronger during the transit through the black hole.

\begin{figure*}
\subfloat[\label{fig:sk1}]{\includegraphics[width=.45\linewidth]{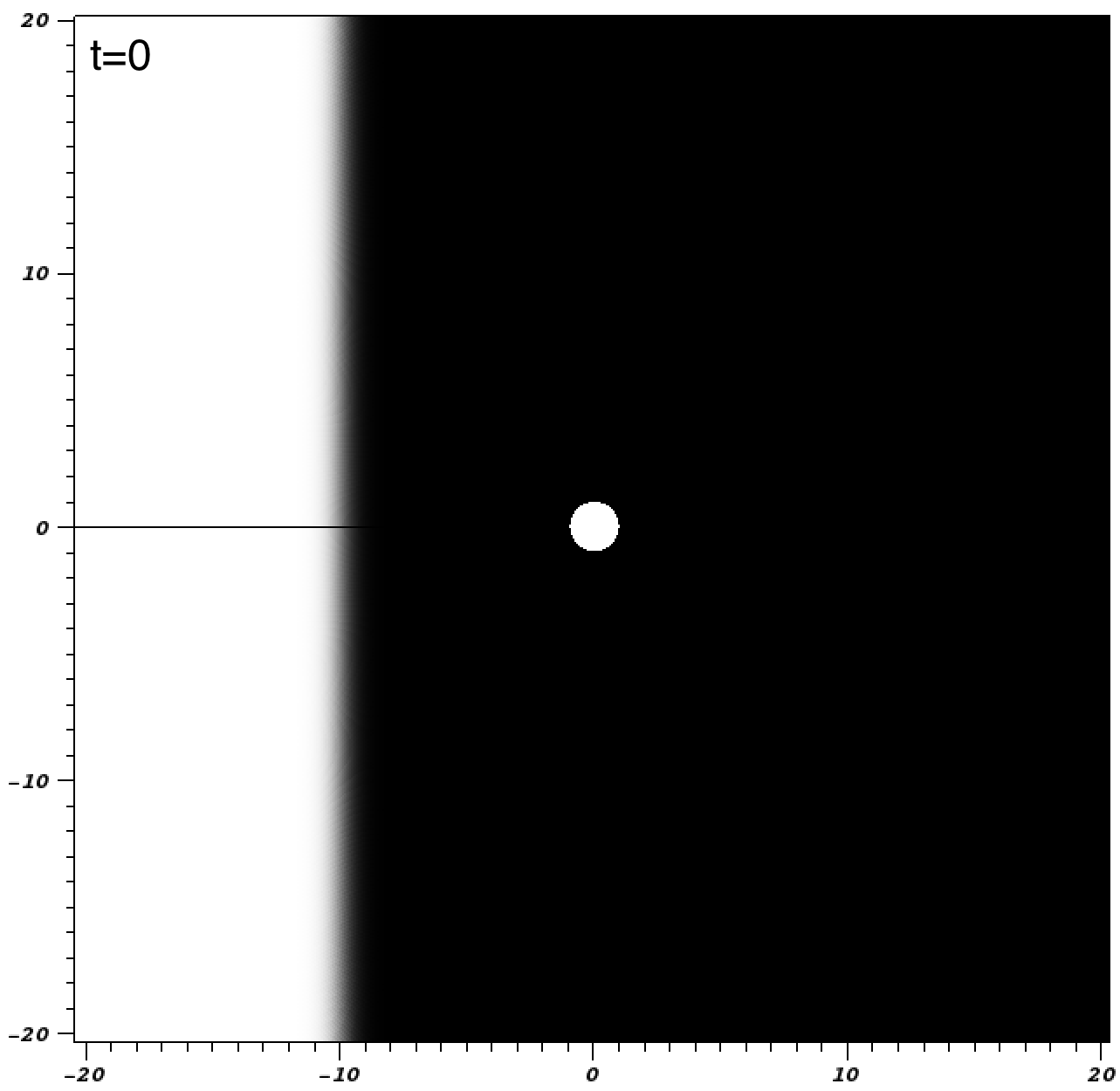}}
\subfloat[\label{fig:sk2}]{\includegraphics[width=.45\linewidth]{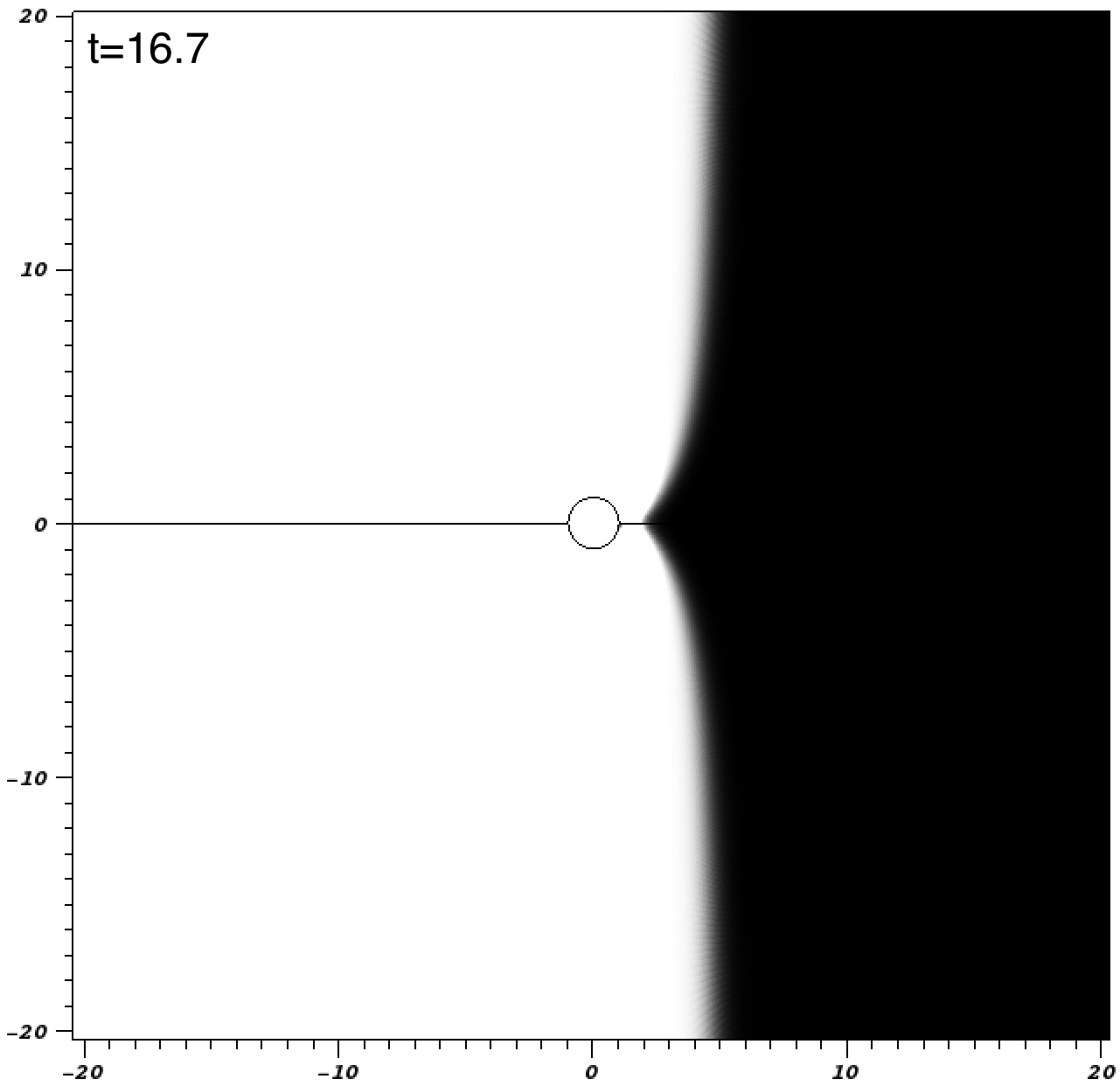}}\hfill
\subfloat[\label{fig:sk3}]{\includegraphics[width=.45\linewidth]{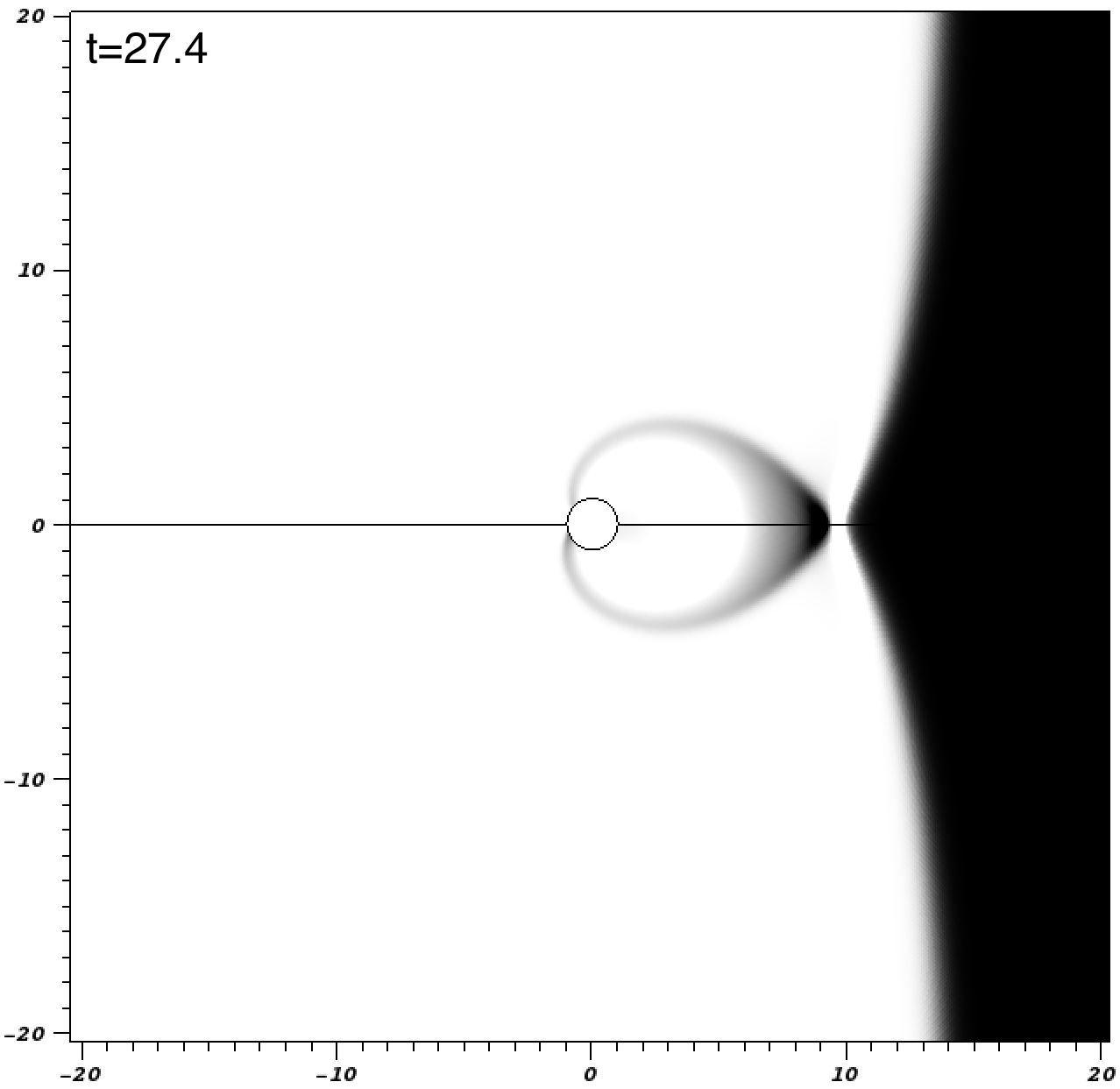}}
\subfloat[\label{fig:sk4}]{\includegraphics[width=.45\linewidth]{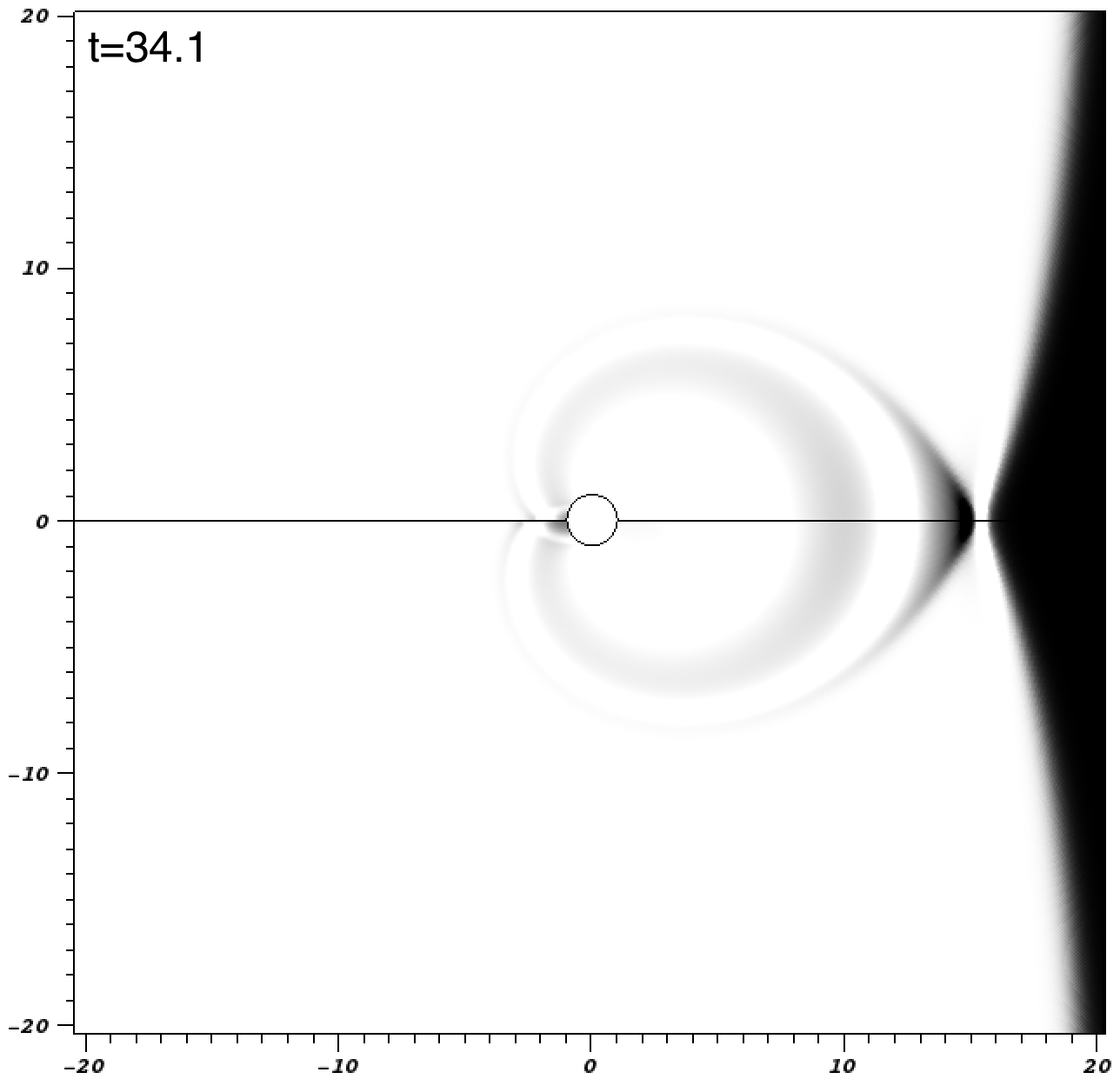}}\hfill
\caption{Successive steps of the evolution of the domain wall with the initial velocity $v=0.9$ near the Schwarzschild black hole of mass $m=1$ (upper halves) and the Kerr black hole with mass $m=1$ and angular momentum $a=0.999$ (lower halves). White and black regions refer to the field in one of two domains.}
\label{fig:schwkerr}
\end{figure*}

We have repeated simulations with the initial velocity $v=0.9$ for a range of values of the black hole angular momentum presented in Table \ref{tab:parameters}. Qualitatively the results were very similar to the ones obtained for the Schwarzschild case. The only differences were relatively small changes in the energy density of the field (increases up to one third when comparing cases with $a=0.999$ and $a=0$) and the tempo of the evolution near the horizon (cf.\ Fig.\ \ref{fig:schwkerr}).

We have also performed long-time simulations, in order to check the stability of domain walls after their passage through a black hole (results are shown in Fig.\ \ref{fig:stab}). It appears that even though the domain wall is initially distorted by the black hole, it eventually returns to its planar shape. Also the ringing modes created during the transit dissipate.

\section{Comparison with the Reissner-Nordstr\"{o}m spacetime}

It is common to consider the Reissner-Nordstr\"{o}m solution as a toy model for the more complicated Kerr geometry \cite{Poi89,Bur97,Daf03,Fic15}. Both spacetimes share similar horizon \cite{Poi90} and causal \cite{Akc11} structures, the former being simpler to deal with due to the spherical symmetry. The Reissner-Nordstr\"{o}m solution can be written in the Kerr-Schild-type coordinates \footnote{In fact the general Kerr-Newman metric can be written in such coordinates; they reduce to Eq.\ (\ref{eqn:Kerr}) for $Q\to 0$ and to Eq.\ (\ref{eqn:RN}) for $a\to 0$.}, in this context usually called Eddington-Finkelstein-type coordinates:
\begin{align} 
ds_{RN}^2=&-\left(1-\frac{2m}{r}+\frac{Q^2}{r^2}\right)dt^2+2\left(\frac{2m}{r}-\frac{Q^2}{r^2}\right)dt dr
\nonumber \\
&+\left(1+\frac{2m}{r}-\frac{Q^2}{r^2}\right)dr^2+r^2 d\theta^2+r^2 \sin^2\theta d\varphi^2,
\label{eqn:RN}
\end{align}
where $Q$ is a charge of the black hole. The horizons in this spacetime occur for the radii $r$ satisfying the condition
\begin{align} 
r^2-2mr+Q^2=0,
\label{eqn:RNhor}
\end{align}
which may be easily compared to the Kerr horizons which correspond to the roots of Eq.\ (\ref{eqn:Delta}). The d'Alembert operator with respect to metric (\ref{eqn:RN}) can be written as
\begin{align} 
\Box_{RN}\phi=&\frac{1}{r^2}\left[-\left(r^2+2mr-Q^2\right) \partial^2_t \phi+2m \partial_t \phi \right.\nonumber\\
&\left.+(4m r-2Q^2)\partial_t\partial_r \phi \right. \nonumber \\
&\left.+\partial_r( (r^2-2mr+Q^2) \partial_r\phi)\right.\nonumber\\
&\left.+\frac{1}{\sin\theta}\partial_\theta(\sin\theta\partial_\theta\phi)+\frac{1}{\sin^2\theta}\partial_\varphi\partial_\varphi\phi\right].
\label{eqn:DeLambertRN}
\end{align}
The energy density of the field is given by
\begin{align} 
\rho=&\frac{1}{2}\left(1+\frac{2m}{r}-\frac{Q^2}{r^2}\right)(\partial_t \phi)^2 \nonumber \\
&+\frac{1}{2}\frac{1+\left(\frac{2m}{r}-\frac{Q^2}{r^2}\right)^2}{1+\frac{2m}{r}-\frac{Q^2}{r^2}}(\partial_r \phi)^2 \nonumber\\
&-\left(\frac{2m}{r}-\frac{Q^2}{r^2}\right)\partial_t\phi \partial_r\phi+\frac{1}{2r^2}(\partial_\theta\phi)^2\nonumber\\
&+\frac{1}{2r^2\sin^2\theta}(\partial_\theta\varphi)^2+V(\phi),
\label{eqn:edensityRN}
\end{align}
which is well defined above the inner horizon. We get the equation of motion in the Reissner-Nordstr\"{o}m spacetime by putting the wave operator (\ref{eqn:DeLambertRN}) into Eq. (\ref{eqn:EomCur}).

\begin{figure*}
\subfloat[\label{fig:krn1}]{\includegraphics[width=.45\linewidth]{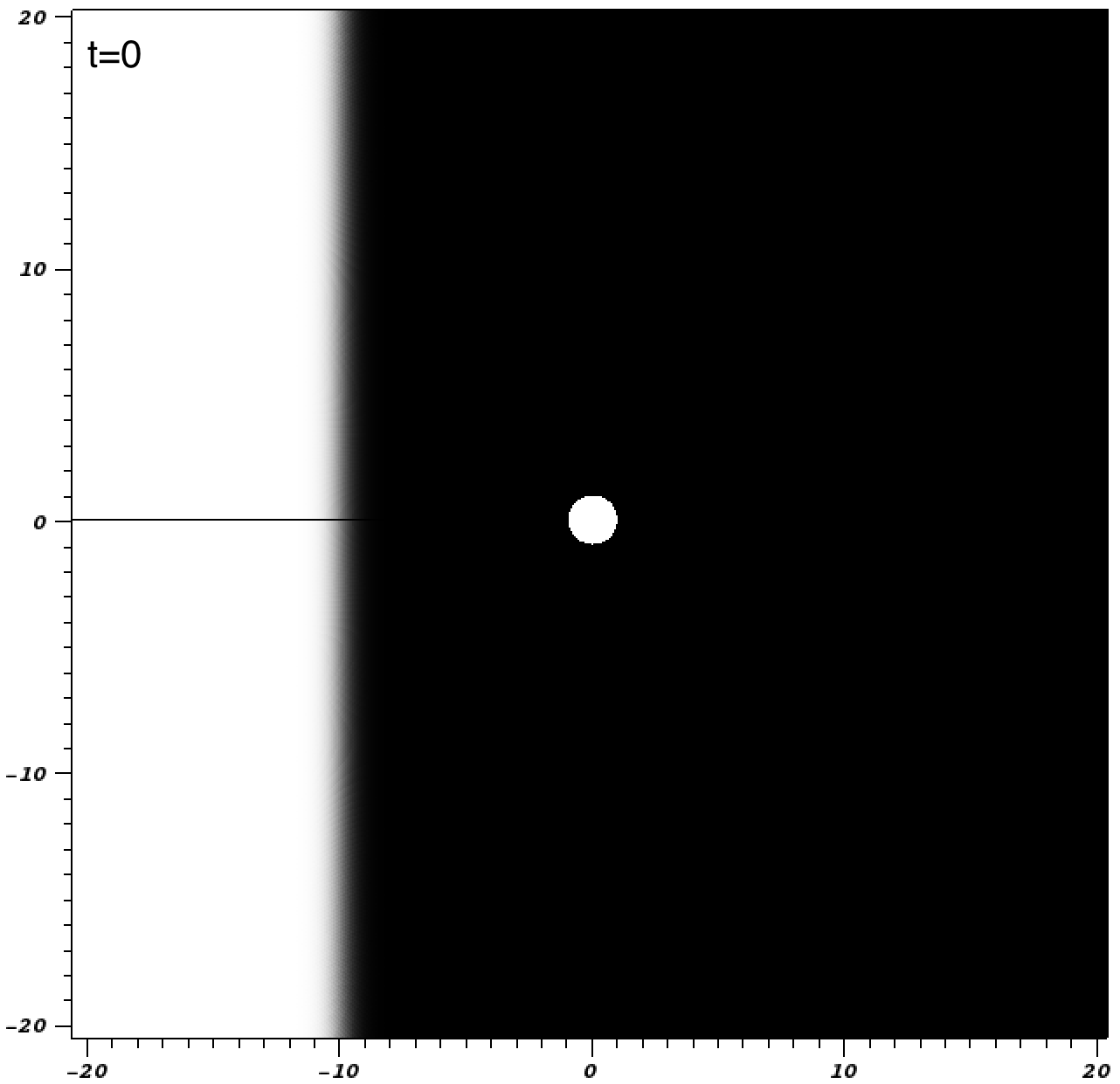}}
\subfloat[\label{fig:krn2}]{\includegraphics[width=.45\linewidth]{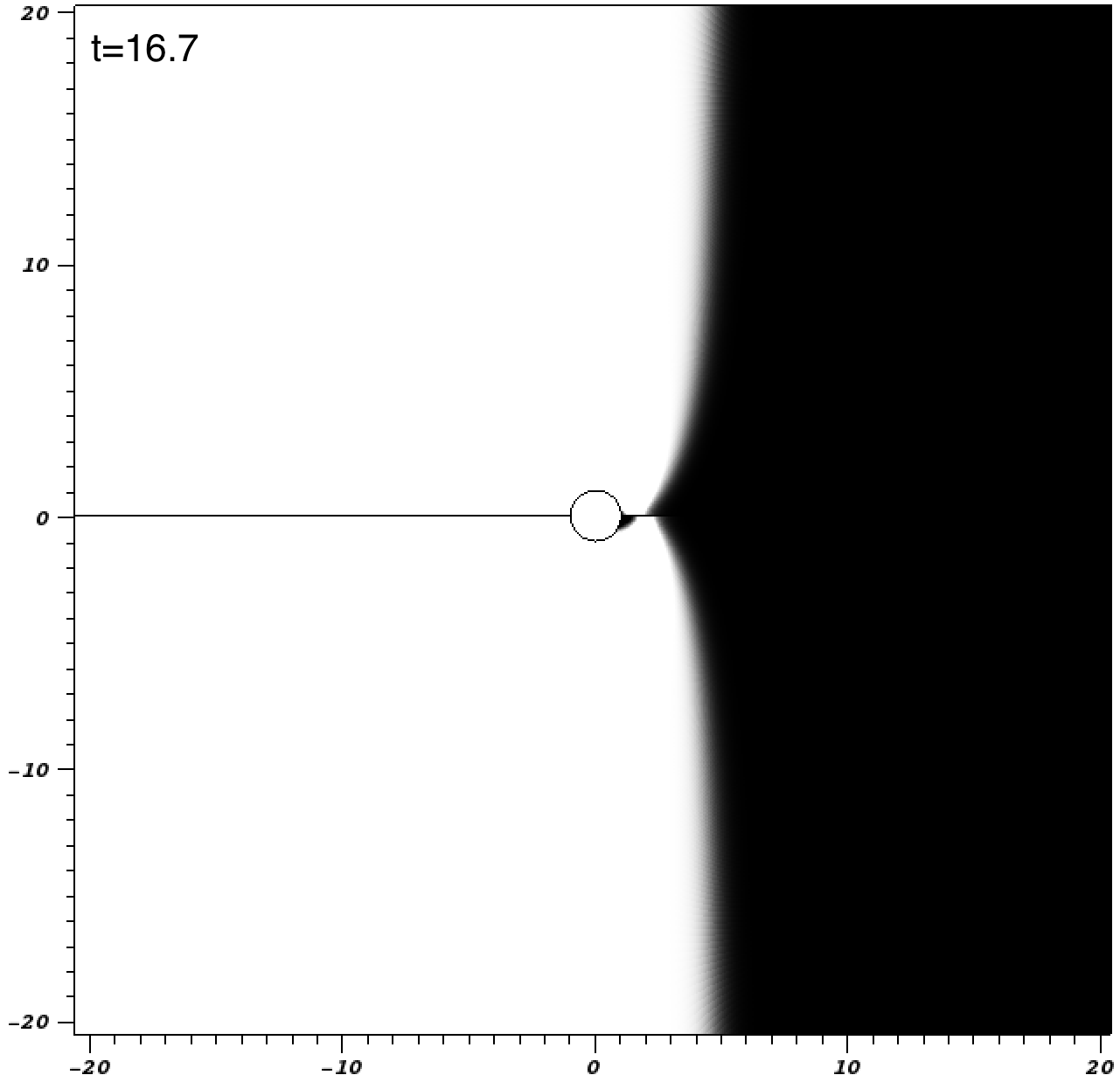}}\hfill
\subfloat[\label{fig:krn3}]{\includegraphics[width=.45\linewidth]{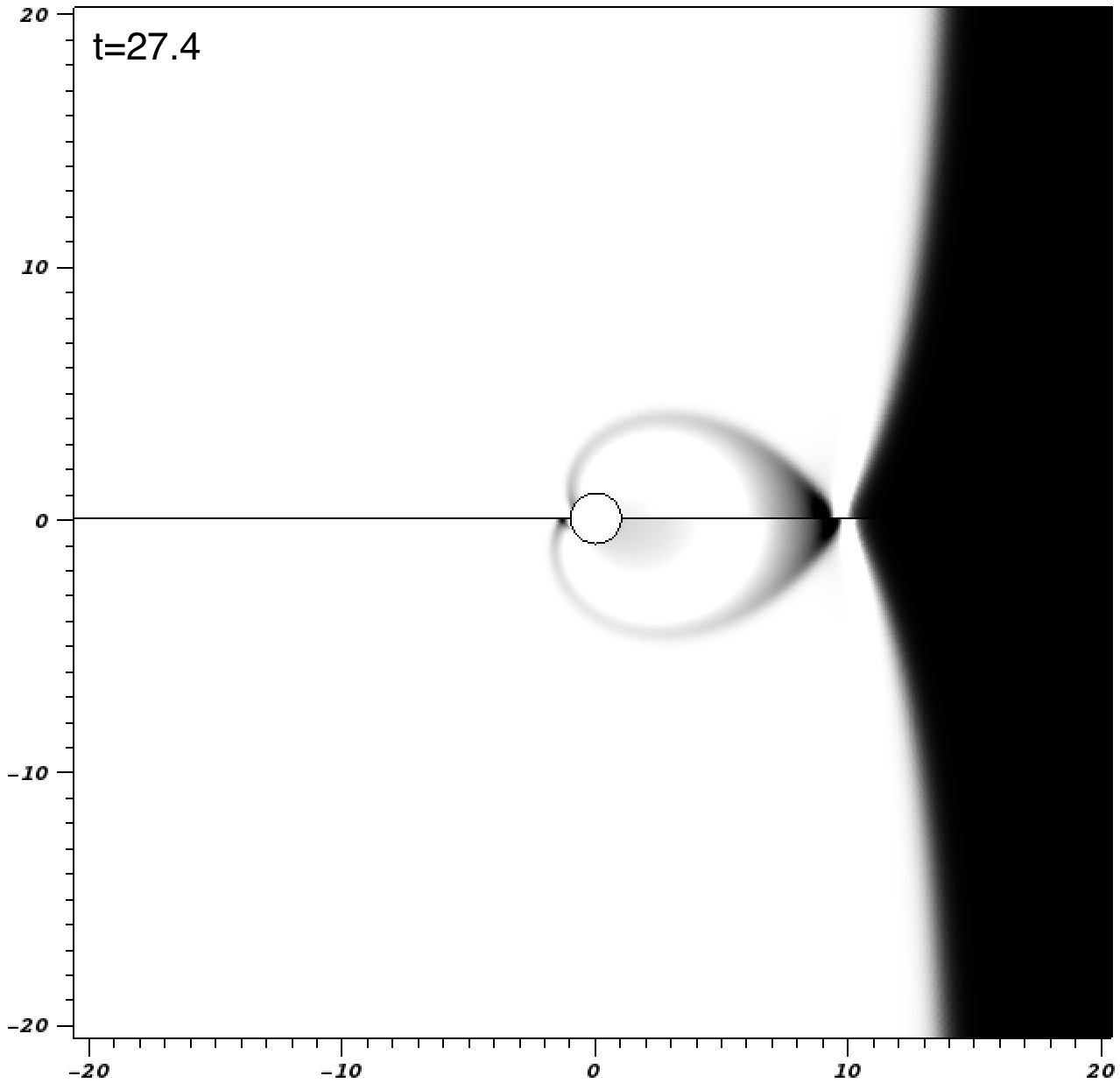}}
\subfloat[\label{fig:krn4}]{\includegraphics[width=.45\linewidth]{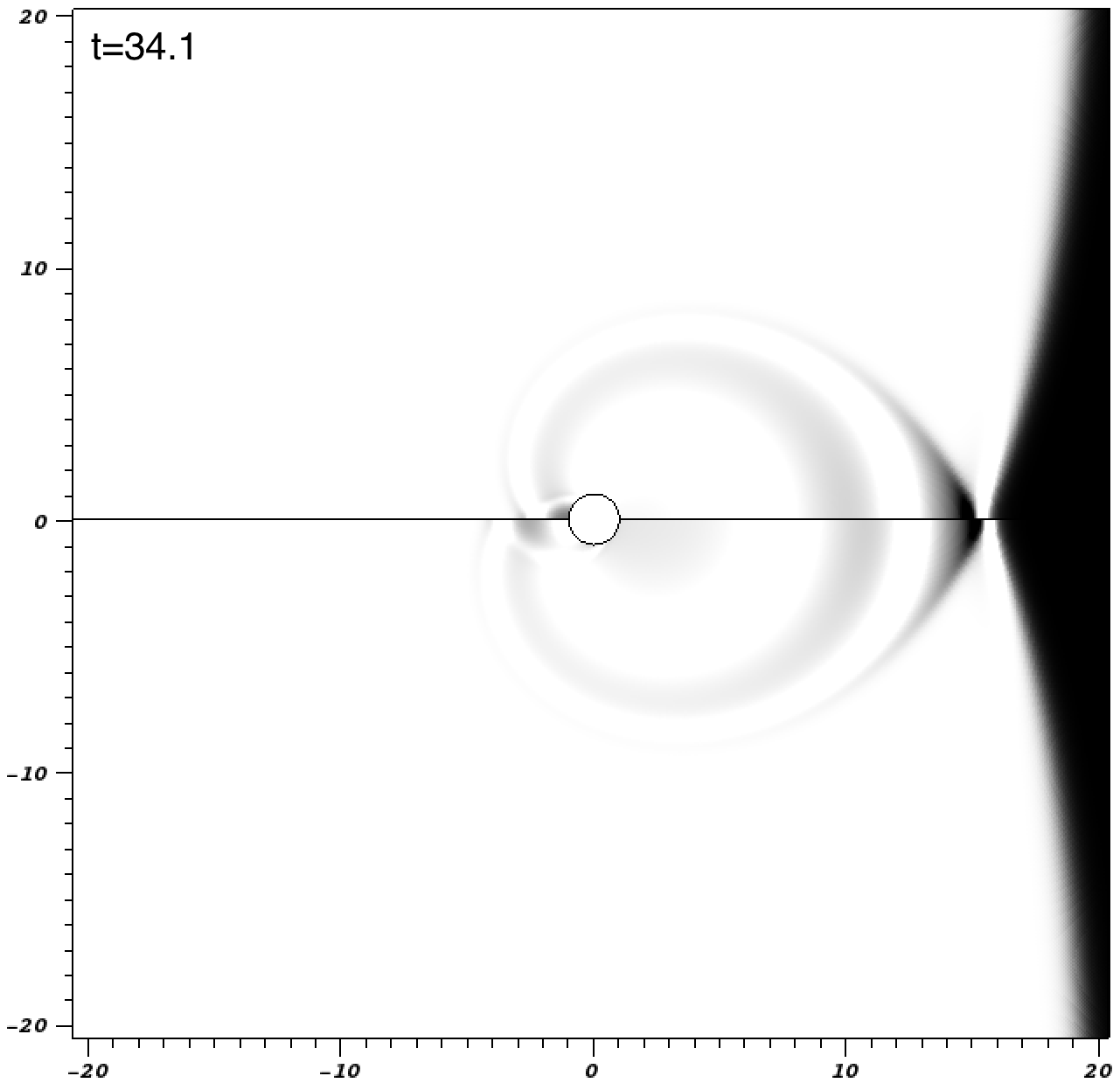}}\hfill
\caption{Successive steps of the evolution of the domain wall with the initial velocity $v=0.9$ near the Kerr black hole with the mass $m=1$ and the angular momentum $a=0.999$ (upper halves) and the Reissner-Nordstr\"{o}m black hole with the mass $m=1$ and the charge $Q=0.999$ (lower halves). White and black regions refer to the field in one of two domains.}
\label{fig:kerrRN}
\end{figure*}

In the following, we compare the transits of domain walls through the Kerr and Reissner-Nordstr\"{o}m black holes. We have performed simulations for Reissner-Nordstr\"{o}m geometries with $m = 1$ and values of charge $Q$ equal to the values of spin $a$ in Table \ref{tab:parameters}. The differences between transit through a charged black hole, and a spinning black hole are rather quantitative than qualitative. The main discrepancy lies in the tempo of evolution in the near-horizon area. As it can be seen in Fig.\ \ref{fig:kerrRN}, the evolution for the Reissner-Nordstr\"{o}m black hole is accelerated in comparison to the evolution obtained for the Kerr geometry. This effect is similar to the one observed in the comparison of transits through Schwarzschild and Kerr black holes (Fig.\ \ref{fig:schwkerr}), but seems to be stronger. Apart from the observed time shift, the shapes of the observed field configurations and the values of the energy density are similar in these two cases.

\section{Summary}

We investigated the behaviour of a low-mass domain wall in the so-called $\phi^4$ scalar field model, after a transit through a Schwarzschild, Kerr or Reissner-Nordstr\"{o}m black hole. The results show that such an event not only distorts the initially planar domain wall, but also creates an additional structure resembling ringing modes in the scalar field. During the latter evolution the domain wall returns to its initial shape, and the ringing structure dissipates. This suggests that domain walls are stable against passages through black holes.

The amplitude of the observed perturbations of the domain wall grows with the increase of its initial velocity. The angular momentum of the black hole has a similar, but weaker impact --- spinning black holes disturb the domain wall slightly stronger than the static ones. Charged black holes affect the domain walls in a similar manner to the spinning ones, except for the accelerated evolution near the black hole in former case. This fact agrees with the practice of using the Reissner-Nordstr\"{o}m geometry as a model for the Kerr spacetime --- the results are qualitatively very similar.

The results presented here show several directions in which investigations of this topic could be conducted further. The most obvious generalisation consists in performing similar simulations for the Kerr spacetime in the three-dimensional setting, without the restriction to the axisymmetry. The other possibility is to consider other scalar field potentials, especially the ones with a richer ground state manifold, such as the sine-Gordon model. One could also focus on the ringing modes appearing during the transit of the domain wall through the black hole, and investigate them in a more systematic way. It is especially interesting to find out if they are the quasinormal modes, or if they arise due to the nonlinearity of the model. We hope to study these issues in the future work.

\begin{acknowledgments}
FF would like to thank Jerzy Knopik and Mieszko Rutkowski for fruitful discussions and helpful ideas, and Victor Flambaum for the inspiration to investigate the subject of this article. This project was supported by the Polish Ministry of Science and Higher Education within the Diamond Grant (grant 0143/DIA/2016/45). PM acknowledges the financial support of the Narodowe Centrum Nauki Grant No.\ DEC-2012/06/A/ST2/00397.
\end{acknowledgments}

\appendix
\section{Iterative Crank-Nicolson method}\label{sec:cn}

The Crank-Nicolson method is an implicit scheme of solving initial value problems for sets of $N$ ordinary differential equations of the form
\begin{align}
\frac{du}{dt} = F(u), \quad u(t = t^0) = u^0.
\end{align}
Here $u \colon [t^0,\infty) \to \mathbb R^N$ is a vector of unknowns, and $F \colon \mathbb R^N \to \mathbb R^N$ is a given function. The time $t$, and the independent variable $u$ are discretised as $t^n = t^0 + n \Delta t$, $u^n = u(t^n)$, $n = 0, 1, \dots$. The method itself is defined by the relation
\begin{align}
u^{n+1}=u^n+\frac{\Delta t}{2}\left[ F(u^{n+1}) + F(u^n) \right].
\label{eqn:CN2}
\end{align}
The so-called iterative Crank-Nicolson scheme is a way of dealing with the implicit term $F(u^{n+1})$. Equations (\ref{eqn:CN2}) are replaced with
\begin{align}
\tilde{u}^{n+1}_0&=u^n,\nonumber \\
\tilde{u}^{n+1}_1 &=u^n+\frac{\Delta t}{2}\left[ F(\tilde{u}^{n+1}_0) + F(u^n) \right],\nonumber \\
\tilde{u}^{n+1}_2 &=u^n+\frac{\Delta t}{2}\left[ F(\tilde{u}^{n+1}_1) + F(u^n) \right],\nonumber\\
&\vdots\nonumber \\
\tilde{u}^{n+1}_k &=u^n+\frac{\Delta t}{2}\left[ F(\tilde{u}^{n+1}_{k-1}) + F(u^n) \right],\nonumber\\
u^{n+1} &=u^n+\frac{\Delta t}{2}\left[ F(\tilde{u}^{n+1}_{k}) + F(u^n) \right],
\label{eqn:CN3}
\end{align}
where we have introduced auxiliary variables $\tilde u_0^{n+1}, \dots, \tilde u_k^{n+1}$. It turns out that already two iterations provide sufficient accuracy \cite{Teu00}. That leads to the formula
\begin{align}
u^{n+1} =u^n+\frac{\Delta t}{2}&\left[ F\left(u^n+\frac{\Delta t}{2}\left[F\left(u^n+F(u^n)\Delta t\right)\right.\right.\right.\nonumber\\
&\left.\left.\left.+F(u^n)\right]\right) + F(u^n) \right].
\label{eqn:CN4}
\end{align}
The scheme given by (\ref{eqn:CN4}) belongs in fact to the class of explicit Runge-Kutta methods and can be written in a more common form
\begin{align}
u^{n+1} = u^n+\Delta t \left(\frac{1}{2} k_1+\frac{1}{2}k_3\right),
\label{eqn:CN5}
\end{align}
where
\begin{align}
k_1 &= F(u^n),\nonumber\\
k_2 &= F(u^n+\Delta t \, k_1),\nonumber\\
k_3 &= F\left(u^n+\Delta t \left(\frac{1}{2}k_1+\frac{1}{2}k_2\right)\right).
\label{eqn:CN6}
\end{align}

\section{Parameters of domain walls}\label{sec:paren}

Since in this work we only consider the evolution of domain walls on fixed Schwarzschild, Kerr, or Reissner-Nordstr\"{o}m backgrounds, the consistency of the model demands ensuring that a rough estimate of the energy of the domain wall contained in the vicinity of the black hole is smaller than the black hole mass.

Equation (\ref{eqn:Energy}) gives the energy of a domain wall in the $1+1$ dimensional flat case (a kink). It is proportional to $\eta^3 \lambda^{1/2} (1-v^2)^{-1/2}$. We will interpret this expression as a surface energy. On the other hand the mass $m$ determines the energy scale and the length scale associated with the black hole. This yields a corresponding energy estimate for the domain wall in the form $\eta^3 \lambda^{1/2} (1-v^2)^{-1/2} m^2$. We require this energy to be much smaller than the mass of the black hole, i.e., 
\begin{equation} 
\frac{\eta^3 \lambda^{1/2}m}{\sqrt{1-v^2}} \ll 1.
\label{eqn:DWBHenergy}
\end{equation}
For $m = 1$ and domain walls with $v < 0.999$, it suffices to set $\eta=0.1$ and $\lambda=100$. Then the left-hand side of Eq.\ (\ref{eqn:DWBHenergy}) is always less than 0.224.

\section{Repulsion from the black hole}\label{sec:rep}

In this Appendix we give a heuristic argument that a domain wall, that is initially initially at rest, is repulsed from the black hole.

For clarity, we will work in Boyer-Lindquist coordinates (denoted in this Appendix as $t$, $r$, $\theta$, $\varphi$). The Kerr metric, written in Boyer-Lindquist coordinates, has the familiar form
\begin{eqnarray}
ds^2 & = & - \left( \frac{\Delta - a^2 \sin^2 \theta}{\Sigma} \right) dt^2 + \frac{\mathcal{A} \sin^2 \theta }{\Sigma} d \varphi^2 + \frac{\Sigma}{\Delta} dr^2 \nonumber \\
&& + \Sigma d \theta^2 - \frac{2 a \sin^2 \theta (r^2 + a^2 - \Delta)}{\Sigma} dt d\varphi,
\label{kerrbl}
\end{eqnarray}
where $\Sigma = r^2 + a^2 \cos^2 \theta$, $\Delta = r^2 + a^2 - 2 m r$, and $\mathcal A = (r^2 + a^2)^2 - \Delta a^2 \sin^2 \theta$.

We consider the following initial data:
\begin{align} 
\phi(t,r,\theta,\varphi)|_{t=0}&=\eta\tanh\left(\eta\sqrt{\frac{\lambda}{2}}(r \cos \theta- z_0)\right), \nonumber \\
\partial_t \phi(t,r,\theta,\varphi)|_{t=0}&=0, \label{eqn:static2}
\end{align}
and assume that $z_0 > 0$. The domain wall described by Eq.\ (\ref{eqn:static2}) is perpendicular to the symmetry axis of the spacetime. Let us consider a point $C$ with coordinates $r = z_0$, $\theta = 0$. It belongs to the intersection of the domain wall and the axis $\theta = 0$. We show that for $t = 0$ the domain wall at point $C$ drifts away from the black hole, i.e., toward larger values of $r$.

Let $\eta^\mu$ be the timelike Killing vector associated with metric (\ref{kerrbl}). We have $\eta^\mu = (1,0,0,0)$. Define the radial momentum $P^r = - T^r_\mu \eta^\mu$, where the energy-momentum is given by Eq.\ (\ref{eqn:TmunuCur}). A simple calculation yields
\[ P^r = - \frac{\Delta}{\Sigma} \partial_t \phi \partial_r \phi. \]
Clearly, outside the black hole, an outgoing wave corresponds to positive $P^r$. Conversely, for the incoming wave one has $P_r < 0$. It is possible to show that $P^r$ is nonnegative at point $C$ for initial data (\ref{eqn:static2}). In practice, it suffices to investigate the sign of $P = - \partial_t \phi \partial_r \phi$. Since the domain wall described by Eq.\ (\ref{eqn:static2}) is initially at rest, we have $P = 0$ for $t = 0$. It is, however, possible to compute the derivative
\begin{align} 
U &= \partial_t P|_{C;t=0} = - \partial_t(\partial_t\phi \; \partial_r \phi)|_{C; t=0} \nonumber \\
&= - \partial^2_t\phi \;  \partial_r \phi|_{C; t=0} - \partial_t\phi\; \partial_t \partial_r\phi|_{C;t=0},
\label{eqn:momentum1}
\end{align}
where the last term vanishes. A rather lengthy calculation involving the equation of motion (\ref{eqn:EomCur}) yields for initial data (\ref{eqn:static2})
\begin{equation} 
U = - \partial^2_t\phi \;  \partial_r \phi|_{C; t=0}  = \frac{m \eta^4 \lambda (a^2 + (z_0 - 2m) z_0)}{(z_0^2 + a^2)^2},
\label{eqn:momentumBL}
\end{equation}
which is positive outside the horizon. Hence, $P$ is nonnegative at $C$, at least for some short period of time $t \in [0,\epsilon)$. We conclude that the domain wall is repulsed from the black hole.

A similar calculation can be also done for the Reissner-Nordstr\"{o}m metric, which in standard coordinates can be written as
\begin{eqnarray}
ds_{RN}^2  & = & - \left( 1 - \frac{2m}{r} + \frac{Q^2}{r^2} \right) dt^2 \nonumber\\
& & + \left( 1 - \frac{2m}{r} + \frac{Q^2}{r^2} \right)^{-1} dr^2 \nonumber \\
& & + r^2 (d\theta^2 + \sin^2 \theta d \varphi^2).
\end{eqnarray}
Repeating the same calculation for the Reissner-Nordstr\"{o}m spacetime we obtain
\begin{equation} 
U = - \partial^2_t\phi \;  \partial_r \phi|_{C; t=0}  = \frac{m \eta^4 \lambda (Q^2 + (z_0 - 2m) z_0)}{z_0^4}.
\end{equation}
Here again, $U$ is positive outside the horizon. This leads to the same conclusion that the black hole drifts away from the black hole.

\end{document}